\documentclass[
aps,
prd,
showpacs,
preprint,
tightenlines,
superscriptaddress,
nofootinbib,
floatfix]
{revtex4}
\usepackage{graphicx,
}
\begin{document}
%
\title{                Three-generation flavor transitions and decays\\
                            of supernova relic neutrinos }
%
%
\author{        G.L.~Fogli}
\affiliation{   Dipartimento di Fisica
                and Sezione INFN di Bari\\
                Via Amendola 173, 70126 Bari, Italy\\}

\author{        E.~Lisi}
\affiliation{   Dipartimento di Fisica
                and Sezione INFN di Bari\\
                Via Amendola 173, 70126 Bari, Italy\\}

\author{        A.~Mirizzi}
\affiliation{   Dipartimento di Fisica
                and Sezione INFN di Bari\\
                Via Amendola 173, 70126 Bari, Italy\\}
\author{        D.~Montanino}
\affiliation{   Dipartimento di Scienza dei Materiali
                and Sezione INFN di Lecce\\
                Via Arnesano, 73100 Lecce, Italy\\}

\begin{abstract}
If neutrinos have mass, they can also decay. Decay lifetimes of
cosmological interest can be probed, in principle, through the
detection of the redshifted, diffuse neutrino flux produced by all
past supernovae---the so-called supernova relic neutrino (SRN)
flux. In this work, we solve the SRN kinetic equations in the
general case of three-generation flavor transitions followed by
invisible (nonradiative) two-body decays. We then use the general
solution to calculate observable SRN spectra in some
representative decay scenarios. It is shown that, in the presence
of decay, the SRN event rate can basically span the whole range
below the current experimental upper bound---a range accessible to
future experimental projects. Radiative SRN decays are also
briefly discussed.
\end{abstract}
\medskip
\pacs{
14.60.Pq, 13.35.Hb, 97.60.Bw, 14.80.Mz} \maketitle

\newcommand{\nuornubar}{{\stackrel{{}_{(-)}}{\nu}\!\!}}
\newcommand{\nubar}{\bar \nu}
\newcommand{\nui}{\nu_i}
\newcommand{\nuj}{\nu_j}
\newcommand{\nubari}{\nubar_i}
\newcommand{\nubarj}{\nubar_j}

\section{Introduction}

 In the last few years, compelling evidence has emerged in
favor of neutrino masses and mixings (see, e.g., \cite{Revi} for
recent reviews). Within a three-generation framework, the known
flavor  eigenstates of neutrinos $\nu_{\alpha}$ and antineutrinos
$\bar\nu_\alpha$ $(\alpha=e,\mu,\tau)$ must then be considered as
linear superpositions of mass eigenstates $\nu_i$ ($i=1,2,3$),
\begin{equation}
\label{U}
\nuornubar_\alpha=\sum_i U_{\alpha i}\nuornubar_i\ ,
\end{equation}
where the mixing matrix $U_{\alpha i}$ (assumed here to be real
for simplicity) can be expressed in terms of three mixing angles
$(\theta_{12},\theta_{13},\theta_{23})$ in the standard
parametrization \cite{Hagi}. The  $\nu_i$ squared mass spectrum
can be cast in the form
\begin{equation}
\label{Mass} (m^2_1,m^2_2,m^2_3)= M^2+\left(-\frac{\delta
m^2}{2},+\frac{\delta m^2}{2},\pm\Delta m^2 \right)\ ,
\end{equation}
where $\delta m^2$ and $\Delta m^2$ govern the two independent
$\nu$ oscillation frequencies, while $M^2$ fixes the absolute
$\nu$ mass scale. Current neutrino oscillation data (see, e.g.,
\cite{Sola,Atmo}) provide the best-fit values:
\begin{eqnarray}
\delta m^2 &\simeq & 7.2\times 10^{-5} \mathrm{\ eV}^2\ ,\label{dm2}\\
|\pm \Delta m^2| &\simeq& 2.0\times 10^{-3} \mathrm{\ eV}^2\ ,\label{Dm2}\\
\sin^2\theta_{12} &\simeq & 0.29 \label{s12}\ ,\\
\sin^2\theta_{23} &\simeq & 0.50 \label{s23}\ ,
\end{eqnarray}
and the upper bound  (at $3\sigma$) \cite{Atmo}
\begin{equation}
\label{s13} \sin^2\theta_{13}<0.067\ .
\end{equation}
The sign of $\Delta m^2$ is currently unknown, while $M^2$ is
bounded from above by laboratory and astrophysical constraints
($\sqrt{M^2}\lesssim {\mathrm{few}}\times 10^{-1}$~eV
\cite{Revi}).

In general, massive neutrinos can not only mix, but also decay
(see, e.g., the reviews in \cite{Dolg,Pakv,Gelm}). Neutrino decay
has been invoked in the past, e.g., to solve the solar neutrino
problem  or the atmospheric neutrino anomaly (see \cite{Pak2} and
references therein). These and other neutrino decay solutions have
not been experimentally validated so far, implying that the decay
lifetimes ($\tau_i$ in the rest frame) must be sufficiently long
to leave unperturbed the current neutrino phenomenology. Roughly
speaking, this can be achieved by imposing $\tau_i E/m_i L\gg 1$
in the laboratory frame, where $L$ and $E$ are the neutrino
pathlength and energy, respectively. Such arguments, properly
refined and applied to solar neutrinos [$L/E \sim O(10^7)$
km/MeV], lead to a safe lower bound \cite{Bell}%
\footnote{This bound can be improved, for $\nu\to\bar\nu$ decays,
by one to two orders of magnitude (depending on specific
scenarios) through the nonobservation of solar $\nu_e\to\bar\nu_e$
transitions in the KamLAND experiment, as recently reported in
Ref.~\cite{KaNu}.}
\begin{equation}
\label{solarbound} \frac{\tau_i}{m_i}\gtrsim 5\times 10^{-4}
\mathrm{\ s/eV}\ .
\end{equation}
Similarly, from the supernova (SN) 1987A neutrino events [$L/E\sim
O(10^{16})$ km/MeV] one might naively guess a much stronger lower
bound,
\begin{equation}
\label{SN87bound} \frac{\tau_i}{m_i}\gtrsim O(10^{5}) \mathrm{\
s/eV}\ ,
\end{equation}
which, however, does not hold for relatively large values of the
mixing angle $\theta_{12}$ \cite{SN87}, as those implied by
current data in Eq.~(\ref{s12}) (see also the comment at the end
of Sec.~IV~C). On the other hand, for neutrino decay effects to be
observable in our universe, it must be roughly $\tau_i E/m_i
\lesssim 1/H_0$, where $H_0$ is the Hubble constant; by setting
$H_0=70 \;\mathrm{km}\,\mathrm{s}^{-1}\,\mathrm{Mpc}^{-1}$ and
taking $E\sim O(10)$~MeV (i.e., in the energy range probed by
supernova neutrinos), a rough upper bound for the ``neutrino decay
observability'' is obtained,
\begin{equation}
\label{upperbound} \frac{\tau_i}{m_i}\lesssim O(10^{11}) \mathrm{\
s/eV}\ .
\end{equation}
The comparison of Eqs.~(\ref{solarbound}) and (\ref{upperbound})
implies that many decades in $\tau_i/m_i$ are still open to
experimental and theoretical investigations.

The exploration of relatively long neutrino lifetimes, and of the
associated decay effects, can be effectively pursued through
astrophysical and cosmological observations. For instance,
evidence for neutrino decay might be signaled by an anomalous
flavor composition of high-energy astrophysical neutrinos
\cite{Be03}. Neutrino decay might also profoundly alter the flux
of supernova relic neutrinos (SRN) and possibly push it close to
the current experimental SRN upper bounds \cite{SKSN,Male}, as
recently pointed out in Ref.~\cite{Ando}. Given the promising
prospects for detecting SRN signals in the Super-Kamiokande (SK)
water Cherenkov detector (if doped with Gd \cite{GADZ}) or in
planned larger detectors such as the Underground Nucleon Decay and
Neutrino Observatory (UNO) \cite{UNNO} or Hyper-Kamiokande
\cite{Hype}, we think that the interesting idea of probing
neutrino decay through SRN data \cite{Ando} appears as an
opportunity which, although challenging, deserves further studies.

On the basis of the previous motivations, in this work we aim at
providing a general framework, as well as specific examples, for
the calculation of nonradiative two-body decays of SRN,
\begin{equation}
\label{Decay}\nu_i\to\nuornubar_j+X\
\end{equation}
(and for analogous antineutrino decays), where $X$ is a
(pseudo)scalar particle, e.g., a Majoron \cite{Gelm}. After
considering the ``standard'' case of $3\nu$ flavor transitions
without decay in Sec.~II, we discuss in Sec.~III the more general
case of $3\nu$ transitions followed by decays, and give the
explicit solution of the neutrino kinetic equations for generic
decay parameters. Specific numerical examples (inspired by
neutrino-Majoron decay models) are given in Sec.~IV, in order to
show representative SRN event rates and energy spectra in the
presence of decay. Conclusions and prospects for future work are
given in Sec.~V.

A final remark is in order. In this paper, we focus on invisible
two-body decays [Eq.~(\ref{Decay})] since they can be tested, in
principle, by future SRN data. Invisible decays into three
neutrinos are not considered here, also because the typical
final-state neutrino energy would generally be too low to be
probed by SRN observations. Radiative two-body neutrino decays,
instead, do not suffer of such ``low-energy'' $\nu$ detection
problem, but are however severely constrained by the current
astrophysical $\gamma$-ray phenomenology. A brief discussion of
the photon background from hypothetical SRN radiative decays is
given in Appendix~A. Appendix~B finally deals with some
subdominant $3\nu$ oscillation effects in the atmospheric neutrino
background, which are often overlooked in SRN searches.

\section{Three-neutrino flavor transitions without decay}

In this Section we consider the standard case of flavor
transitions among three stable neutrinos. Although the no-decay
case for SRN has already been treated in previous works (see,
e.g., \cite{An03,AnSa,Fuku,Ka03,An04} for recent contributions),
we prefer to discuss its calculations in some detail, both to make
the paper self-consistent, and to better appreciate the
differences with the case of unstable neutrinos. In this Section
we also define the notation, as well as our default choices and
approximations for several inputs (from cosmology and
astrophysics, neutrino oscillation data, and supernova
simulations) needed to compute the SRN signal. The same inputs
will be used in the presence of $\nu$ decay (Sec.~IV).

We do not address here the problem of the uncertainties affecting
these---and other equally admissible---input choices, being mainly
interested to highlight the additional effects of neutrino decay
(as compared with the standard case). Careful considerations about
central values and uncertainties of the various inputs will be
necessary, however, when supernova relic neutrinos will be
eventually detected by experiments and used to constrain models,
with or without neutrino decay (see \cite{An04} for a recent
approach using simulated data).

Finally, we remind that three-flavor oscillations affect not only
the signal in SRN searches, but also the background. We discuss in
Appendix~B some subtle $3\nu$ effects which are often neglected in
calculating the atmospheric neutrino background rate.

\subsection{Input from cosmology and astrophysics}

The local (redshift $z=0$) number density  of mass eigenstates
$\nu_i$ per unit of energy, coming from all past core-collapse
supernovae, is given by (see, e.g., \cite{An03}):
\begin{equation}
\label{ni} n_{\nu_i}(E) = \int_0^\infty {dz\, H^{-1}(z)\,
R_{\mathrm{SN}}(z)\, Y_{\nu_i}(E(1+z))}\ ,
\end{equation}
where $R_{\mathrm{SN}}(z)$ is the supernova formation rate (per
unit of time and of comoving volume at redshift $z$),
$Y_{\nu_i}(q)$ is the average yield%
\footnote{The yield of a $\nu$ species represents the integral of
the $\nu$ luminosity over the emission time, and is equal to the
total $\nu$ number times the normalized $\nu$ energy spectrum.
Since SRN signals are intrinsically time-averaged, we use $\nu$
yields $Y_\nu$ (rather than luminosities) throughout this work.
Expressions for the $Y_\nu$ functions  are given below.}
of $\nu_i$ for a typical supernova [per unit of initial,
unredshifted energy $q=E(1+z)$], and $H(z)$ is the Hubble constant
at redshift $z$ which, in standard notation \cite{Hagi}, reads
\begin{equation}
\label{Hz} H(z)=H_0[(1+z)^2(1+\Omega_M z)-\Omega_\Lambda
z(2+z)]^\frac{1}{2}\ .
\end{equation}
Since supernova relic neutrinos are ultrarelativistic ($v\simeq
c$), the above number density per unit of energy $n_{\nu_i}(E)$
can be identified, in natural units ($c=1$), with the local relic
$\nu_i$ flux per unit of time, area, and energy.

We assume, following \cite{An03,Fuku}, that the function
$R_\mathrm{SN}$ is given by
\begin{equation}
\label{RSN} R_{\mathrm{SN}}(z)=0.0122\,
R_{\mathrm{SF}}(z)/M_{\odot}\ ,
\end{equation}
where the star formation rate $R_{\mathrm{SF}}$ is parametrized as
\cite{Mada}
\begin{equation}
\label{RSF}
R_{\mathrm{SF}}(z)=0.3\,h_{65}\,\frac{\exp(3.4z)}{45+\exp(3.8z)}\;\;
M_{\odot}\mathrm{yr}^{-1}\mathrm{Mpc}^{-3}
\end{equation}
for $z<5$ (and $R_{\mathrm{SF}}=0$ otherwise \cite{An03}), with
$h_{65}=H_0/65 \;\mathrm{km}\,\mathrm{s}^{-1}\,\mathrm{Mpc}^{-1}$.
(Other choices are also possible for $R_\mathrm{SF}$; see, e.g.,
\cite{Ka03}.)

The above expression for $R_{\mathrm{SF}}$ actually holds for an
Einstein-de Sitter cosmology $(\Omega_M,\Omega_\Lambda)=(1,0)$.
For a different cosmology one has to apply a correction factor
\cite{An03}
\begin{equation}
\label{RSFnew} R_{\mathrm{SF}} \to R_{\mathrm{SF}}
\frac{[(1+z)^2(1+\Omega_M z)-\Omega_\Lambda
z(2+z)]^\frac{1}{2}}{(1+z)^\frac{3}{2}}\ ,
\end{equation}
which cancels the dependence of $n_{\nu_i}$ upon
$(\Omega_M,\Omega_\Lambda)$ in Eq.~(\ref{ni}). This cancellation
does not apply in the presence of neutrino decay (see Sec.~III~B).
When needed, we shall then fix $\Omega_M=0.3$,
$\Omega_\Lambda=0.7$, and $h_{65}=1.077$ (i.e., $H_0=70
\;\mathrm{km}\,\mathrm{s}^{-1}\,\mathrm{Mpc}^{-1}$), consistently
with recent determinations of cosmological parameters \cite{SDSS}.

\subsection{Input from neutrino oscillation phenomenology}

The smallness of the $\delta m^2/\Delta m^2$ ratio [see
Eqs.~(\ref{dm2}) and (\ref{Dm2})] and of $\sin^2\theta_{13}$ [see
Eq.~(\ref{s13})] lead to the approximate decoupling of the
supernova $3\nu$ dynamics into
 ``high''  ($H$) and ``low'' ($L$) $2\nu$ subsystems, governed by
$(k_H,\sin^2\theta_{13})$ and $(k_L,\sin^2\theta_{12})$,
respectively, where $k_H=\pm \Delta m^2/2E$ and $k_L=\delta
m^2/2E$ are the two independent neutrino wavenumbers (see, e.g.,
\cite{SNan} and references therein). Matter effects \cite{Matt}
are governed by the potential $V=\pm\sqrt{2}\,G_F\,N_e(x)$, where
$N_e(x)$ is the electron density at radius $x$ (roughly falling as
$x^{-3}$ outside the supernova neutrinosphere, up to shock-wave
effects \cite{Shoc}), and the plus (minus) sign refers to
neutrinos (antineutrinos). The four cases corresponding to
$\mathrm{sign}(\Delta m^2)=\pm1$ (normal or inverted hierarchy)
and $\mathrm{sign}(V)=\pm 1$ (neutrinos or antineutrinos) lead, in
general, to different physics.

In supernova neutrino oscillations, it is customary to average out
unobservable interference phases from the beginning, and to work
directly in terms of level crossing probabilities $P_{ij}$
\cite{KuPa}. At the exit from the supernova, the yields of
neutrino mass eigenstates $Y_{\nu_i}$  are then given by
\begin{equation}
\label{Lum1} Y_{\nu_i}=\sum_{\alpha,j}P_{ij}\,|U^{m}_{\alpha
j}|^2\,Y_{\nu_\alpha}\
\end{equation}
(and similarly for antineutrinos), where $U^{m}_{\alpha j}$ and
$Y_{\nu_\alpha}$ are, respectively, the mixing matrix elements in
matter and the neutrino flavor yields at the neutrinosphere.

The usual assumption $Y_{\nu_\mu}=Y_{\nu_\tau}=Y_{\bar
\nu_\mu}=Y_{\bar \nu_\tau}\equiv Y_{\nu_x}$ implies that the inner
matrix product in Eq.~(\ref{Lum1}) depends only on $Y_{\nu_x}$,
$Y_{\nu_e}$ for neutrinos ($Y_{\nu_x}$, $Y_{\bar\nu_e}$ for
antineutrinos), and on the squared mixing matrix elements $|U^m_{e
i}|^2$. Such elements are given by
\begin{equation}
\label{Ue} (|U^m_{e1}|^2,|U^m_{e2}|^2,|U^m_{e3}|^2)=
(\cos^2\theta_{13}^m\,\cos^2\theta_{12}^m,\,\cos^2\theta_{13}^m\,\sin^2\theta_{12}^m,\,
\sin^2\theta_{13}^m)\ ,
\end{equation}
where
\begin{eqnarray}
\cos2\theta_{13}^m&\simeq &\frac{\cos2\theta_{13}-V/k_H}
{\sqrt{(\cos2\theta_{13}-V/k_H)^2+\sin^22\theta_{13}}}\simeq
-\mathrm{sign}(V)\mathrm{sign}(\Delta m^2)\ ,\label{c13m}\\
\cos2\theta_{12}^m&\simeq &\frac{\cos2\theta_{12}-V/k_L}
{\sqrt{(\cos2\theta_{12}-V/k_L)^2+\sin^22\theta_{12}}} \simeq
-\mathrm{sign}(V)\ ,\label{c12m}
\end{eqnarray}
and we have used the fact that $|V/k_{H,L}|\gg 1$ at the
neutrinosphere.

A further simplification of Eq.~(\ref{Lum1}) comes from the
adiabaticity of the $1\leftrightarrow 2$ level crossing in the $L$
subsystem [$P_{L}\simeq0$ for the mass-mixing values in
Eqs.~(\ref{dm2}) and (\ref{s12})]. The matrix $P_{ij}$ has then
(at most) one nontrivial $2\times 2$ submatrix with indices
$(k,l)$ and off-diagonal entry $P_H$,  where $P_H$ is the
$k\leftrightarrow l$ crossing probability in the $H$ subsystem
(equal for $\nu$ and $\bar \nu$ \cite{SNan}). The final results
for the yield of the $i$-th mass eigenstate at the exit from the
supernova are collected in Table~I.
\begin{table}[t]
\caption{\footnotesize\baselineskip=4mm Relevant mixing matrix
elements and supernova neutrino yields in the four possible $3\nu$
scenarios. The $k\leftrightarrow l$ level crossing in the $H$
system (if any) is reported in the 5th column. See the text for
details. \smallskip}
\begin{ruledtabular}
\begin{tabular}{l|cccc|l}
 Scenario &  $|U^m_{e1}|^2$ & $|U^m_{e2}|^2$ &
$|U^m_{e3}|^2$ & ~~~$k\leftrightarrow l$~~~ & Yield of
$i$-th $\nu$ state\\
\hline
&&&&&$Y_{\nu_1}=Y_{\nu_x}$\\
$\nu$, normal hierarchy & 0 & 0 & 1 & $3\leftrightarrow 2$ &
$Y_{\nu_2}=(1-P_H)Y_{\nu_x}+P_H Y_{\nu_e}$\\
&&&&&$Y_{\nu_3}=P_H Y_{\nu_x}+(1-P_H)Y_{\nu_e}$\\
\hline
&&&&&$Y_{\bar\nu_1}=Y_{\bar\nu_e}$\\
$\bar\nu$, normal hierarchy & 1 & 0 & 0 & --- &
$Y_{\bar\nu_2}=Y_{\nu_x}$\\
&&&&&$Y_{\bar\nu_3}= Y_{\nu_x}$\\
\hline
&&&&&$Y_{\nu_1}= Y_{\nu_x}$\\
$\nu$, inverted hierarchy~~ & 0 & 1 & 0 & --- &
$Y_{\nu_2}= Y_{\nu_e}$\\
&&&&&$Y_{\nu_3}= Y_{\nu_x}$\\
\hline
&&&&&$Y_{\bar\nu_1}=(1-P_H)Y_{\nu_x}+P_H Y_{\bar\nu_e}$\\
$\bar\nu$, inverted hierarchy & 0 & 0 & 1 & $3\leftrightarrow 1$ &
$Y_{\bar\nu_2}=Y_{\nu_x}$\\
&&&&&$Y_{\bar\nu_3}=P_H Y_{\nu_x}+(1-P_H)Y_{\bar\nu_e}$\\
\end{tabular}
\end{ruledtabular}
\end{table}

In general, the crossing probability $P_H$ depends on the $\nu$
potential profile $V(x)$ (see, e.g., \cite{SNan}). This dependence
vanishes in the limiting cases of adiabatic transitions
($P_H\simeq 0$) and strongly nonadiabatic transitions ($P_H\simeq
1$), which correspond roughly to $\sin^2\theta_{13}\gg 10^{-4}$
and $\sin^2\theta_{13}\ll 10^{-5}$, respectively (see, e.g.,
\cite{Luna}). In our numerical examples, we shall consider for
simplicity only the two limiting cases,%
\footnote{In general, $P_H$ is a function of energy \cite{Luna},
and possibly of time \cite{SNan}. However, energy- and
time-dependent effects on SRN signals are relatively small (as
compared with the decay effects that we shall discuss) and are not
considered in this work.}
\begin{equation}
\label{1or0} P_H=0\mathrm{\ or\ }1\ .
\end{equation}

For our purposes, we can also neglect the secondary corrections
due to Earth matter crossing \cite{SNan,Luna} for arrival neutrino
directions below the horizon. In this approximation, the number
density of SRN in the flavor basis is simply given by
\begin{equation}
n_{\nu_\alpha}(E) = \sum_{i} |U_{\alpha i}|^2 n_{\nu_i}(E)\ ,
\end{equation}
and similarly for antineutrinos. Finally, in the
phenomenologically interesting case of supernova relic $\bar\nu_e$
one gets, up to negligible terms of $O(\sin^2{\theta_{13}})$,
\begin{equation}
\label{nubare} n_{\bar\nu_e}(E) \simeq \cos^2\theta_{12}\,
n_{\bar\nu_1}(E)+\sin^2\theta_{12}\, n_{\bar\nu_2}(E)\ ,
\end{equation}
where $\sin^2\theta_{12}$ is taken from Eq.~(\ref{s12}).

\subsection{Input from supernova simulations}

Concerning the three relevant $\nu$ flavor yields at the
neutrinosphere $Y_{\nu}$ ($\nu=\nu_e,\bar\nu_e,\nu_x$), we assume
for simplicity equipartition of the total binding energy $E_b$
(taken equal to $3\times 10^{53}$ erg) among all flavors,
\begin{equation}
\label{Ybeta} Y_\nu(E)= \frac{E_b}{6 \langle
E\rangle_\nu}\,\varphi(E;\langle E\rangle_\nu,\alpha)\ ,
\end{equation}
where $\varphi(E)$ is the normalized neutrino spectrum ($\int dE
\varphi=1$), $\langle E\rangle_\nu$ is the average $\nu$ energy,
and $\alpha$ is a spectral parameter. A useful spectral
parametrization is given in Ref.~\cite{Spec}:
\begin{equation}
\label{varphi} \varphi(E)=\frac{(\alpha+1)^{(\alpha+1)}}{
\Gamma(\alpha+1)}\left(\frac{E}{\langle E
\rangle_\nu}\right)^\alpha \frac{e^{-(\alpha+1)E/\langle
E\rangle_\nu}}{\langle E \rangle_\nu}\ ,
\end{equation}
where $\Gamma$ is the Euler gamma function. From the admissible
parameter ranges in Ref.~\cite{Spec}, we pick up the following
values:
\begin{equation}
\label{Eave} (\langle E \rangle_{\nu_e},\langle E
\rangle_{\bar\nu_e},\langle E \rangle_{\nu_x})
 = (12,15,18)\ \mathrm{MeV}\ ,
\end{equation}
\begin{equation}
\label{alpha}\alpha=3\ .
\end{equation}

Figure~1 shows the functions $Y_{\nu_e}(E)$, $Y_{\bar \nu_e}(E)$,
and $Y_{\nu_x}(E)$, which follow from the previous assumptions. We
emphasize that such assumptions are not meant to provide
``reference'' or ``best'' neutrino spectra, but just a reasonable
input for our numerical computations and for the comparison of
results with and without decay.

\begin{figure}[t]
\vspace*{+0.0cm}\hspace*{-0.2cm}
\includegraphics[scale=0.38]{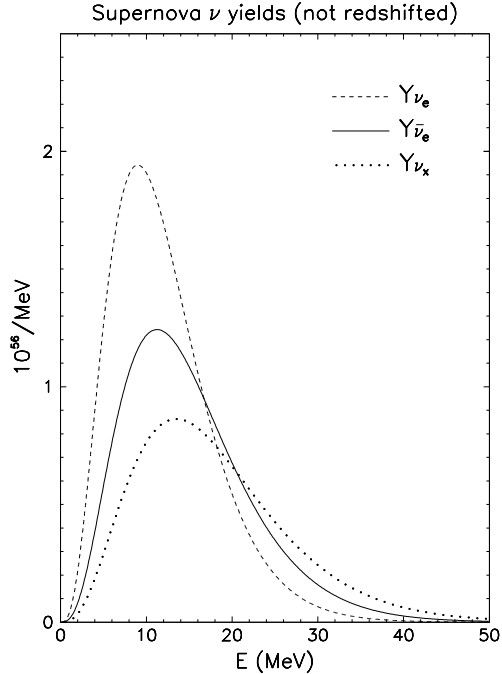}
\vspace*{-0.cm} \caption{\label{fig1}
\footnotesize\baselineskip=4mm  Supernova neutrino yields of
$\nu_e$, $\bar\nu_e$, and $\nu_x$ (curves labelled by $Y_{\nu_e}$,
$Y_{\bar\nu_e}$, $Y_{\nu_x}$, respectively), as defined through
Eqs.~(\ref{Ybeta})--(\ref{alpha}). }
\end{figure}

\subsection{Results for supernova relic $\bar\nu_e$ fluxes and positron spectra (no decay)}

Supernova relic $\bar\nu_e$'s can produce observable signals
(positrons) through the reaction:
\begin{equation}
\bar\nu_e + p \to n + e^+\ .
\end{equation}
For our purposes, the cross section for this process can be
approximated  at zeroth order  \cite{Xsec}, with
$E_{e^+}=E_\nu-1.293$~MeV, where $E_{e^+}$ is the total (true)
positron energy. We will generically refer to ``water targets''
but not to specific detectors; therefore, we will not include
experiment-dependent details such as the difference between true
and measured positron energy (i.e., the detector resolution
function), or the detector efficiency function.

\begin{figure}
\vspace*{-0.0cm}\hspace*{-0.2cm}
\includegraphics[scale=0.95]{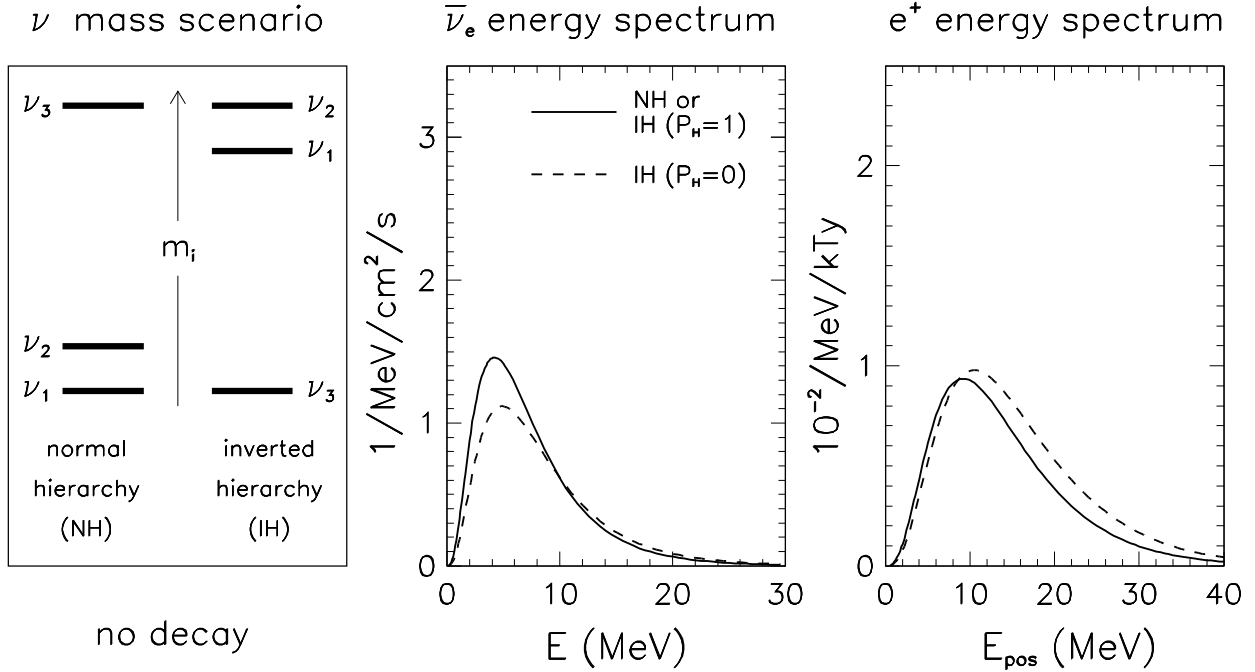}
\vspace*{-0.7cm} \caption{\label{fig2}
\footnotesize\baselineskip=4mm Supernova relic $\bar\nu_e$
spectrum (middle panel), and associated positron spectrum from
$\bar\nu_e+p\to n+e^+$ (right panel), in the presence of $3\nu$
flavor transitions without decay. The spectra depend on the mass
hierarchy (left panel), which can be either normal (NH) or
inverted (IH). In the latter case, there is a further dependence
on the crossing probability $P_H$. The representative cases
$P_H=0$ (``large $\theta_{13}$'') and $P_H=1$ (``small
$\theta_{13}$'') are shown.}
\end{figure}

Figure~2 shows our SRN flux calculations for no decay, in both
cases of normal hierarchy (NH) and inverted hierarchy (IH), as
reported in the left panel. The middle panel refers to the
absolute supernova relic $\bar\nu_e$ fluxes [Eq.~(\ref{nubare})]
in units of $\bar\nu_e$/MeV/cm$^2$/s, as a function of neutrino
energy. The right panel refers to the absolute positron event rate
per MeV and per kton-year (kTy) in water, as a function of the
true positron energy. In the case of normal hierarchy (solid
curves), the $\bar\nu_e$ and positron spectra do not depend on the
crossing probability $P_H$ (see Table~I). This case is
indistinguishable from the case of inverted hierarchy with $P_H=1$
\cite{AnSa}. In both such cases (NH with any $P_H$,  or IH with
$P_H=1$) it is
\begin{equation}
\label{ni2} n_{\bar\nu_e}(E) = \int_0^\infty {dz\, H^{-1}(z)\,
R_{\mathrm{SN}}(z)\, [\cos^2\theta_{12}\,Y_{\bar\nu_e}(E(1+z))
+\sin^2\theta_{12}Y_{\nu_x}(E(1+z))]}\ .
\end{equation}
In the case of inverted hierarchy with $P_H=0$ (dashed curves in
Fig.~2), the supernova relic $\bar\nu_e$ spectrum depends only on
$Y_{\nu_x}$,
\begin{equation}
\label{ni3} n_{\bar\nu_e}(E) = \int_0^\infty {dz\, H^{-1}(z)\,
R_{\mathrm{SN}}(z)\, Y_{\nu_x}(E(1+z))}\ ,
\end{equation}
and is thus peaked at somewhat higher energy [see
Eq.~(\ref{Ybeta}) and Fig.~1].

The results in Fig.~2 are in agreement with previous absolute
estimates of the neutrino and positron spectra in water-Cherenkov
detectors (see, e.g., \cite{AnSa}), modulo different choices for
(some of) the inputs of SRN calculations. As we shall see, such
results can be profoundly modified by neutrino decay.

\section{Three-neutrino flavor transitions and decays}

In this Section we discuss and solve the neutrino kinetic
equations in the general case of $3\nu$ flavor transitions plus
decay. We consider generic rest-frame
(anti)neutrino lifetimes $\tau_i$ and associated decay widths,%
\footnote{In the laboratory frame, the widths are multiplied by
the standard relativistic factor $m_i/E$.}
\begin{equation}
\label{tau}
\tau_i^{-1}=\Gamma_i=\sum_{m_j<m_i}\Gamma(\nui\to\nuj)\
+\Gamma(\nui\to\nubarj)\ .
\end{equation}
Also, in this Section we do not set restrictions on branching
ratios%
\footnote{In the following, we shall often use the symbol
``$\nu$'' loosely, to indicate both neutrinos and antineutrinos. A
distinction between $\nu$ and $\bar\nu$ will be made when needed
to avoid ambiguities.}
\begin{equation}
\label{Branch} B(\nu_i\to\nu_j)=\Gamma(\nu_i\to\nu_j)/\Gamma_i\ ,
\end{equation}
and on neutrino decay energy spectra
\begin{equation}
\label{psi}
\psi_{\nu_i\to\nu_j}(E_i,E_j)=\mathrm{Prob}[\nu_i(E_i)\to\nu_j(E_j)]\
\end{equation}
(normalized to unity, $\int dE_j\, \psi(E_i,E_j)=1$).

Notice that, for $\tau_i/m_i$ values above the bound in
Eq.~(\ref{solarbound}), supernova relic $\nu$ flavor transitions
occur in matter (and become incoherent) well before neutrino decay
losses become significant, so that hypothetical interference
effects between the two phenomena \cite{Lind} can be neglected.
For our purposes, flavor transitions inside the supernova can thus
be taken as decoupled from the subsequent (incoherent) propagation
and decay of mass eigenstates in vacuum.

An important remark is in order. In this work we consider only
{\em vacuum\/} neutrino decays. It is possible, however, to
construct models where fast invisible decays can be triggered by
matter effects \cite{Bere,GLam} at the very high densities
characterizing the supernova neutrinosphere, even in the absence
of vacuum decays (e.g, for diagonal neutrino-Majoron couplings;
see, e.g. \cite{Kach}). In such scenarios, matter-induced decays
might thus occur {\em before\/} flavor transitions in supernovae,
leading to a phenomenology rather different from the one
considered in this paper (and subject to model-dependent
constraints from supernova energetics
\cite{Kach,Smir,Choi,Toma,Farz}). We emphasize that the results
discussed in the following sections are generally applicable to
vacuum neutrino decays occurring after flavor transitions, while
possible matter-induced fast decays are beyond the scope of this
work.

\subsection{Kinetic equations}

In the presence of $\nu$ decay, the local ($z=0$) SRN number
density per unit energy $n_{\nu_i}(E)$ depends on the history of
all past ($z>0$) decays having the $\nu_i$  as initial or final
mass state. The functions $n_{\nu_i}(E,z)$ (number of $\nu_i$ per
unit of comoving volume and of energy at redshift $z$) can be
obtained through a direct integration of the neutrino kinetic
equations, as described below.

The generic form of the kinetic equations for the phase-space
distribution function $f$ is
\begin{equation}
\label{Kinetic} {\cal L}[f]={\cal C}[f]\ ,
\end{equation}
where $\cal L$ is the Liouville operator and $\cal C$ is the
collision operator, embedding creation and destruction terms for
the particle described by $f$ (see, e.g., \cite{Dolg,Kolb}).

For ultrarelativistic relic neutrinos $\nu_i$ $(E\simeq
p=|\overrightarrow{p}|)$, described at time $t$ by
$n_{\nu_i}(E,t)=4\pi p^2 f R^3(t)/R_0^3 $, the Liouville operator
takes
the form%
\footnote{$R(t)$ is the universe scale factor for a
Friedmann-Robertson-Walker metric, with $H(t)=\dot{R}(t)/R(t)$ in
standard notation \cite{Hagi}. With respect to Ref.~\cite{Kolb},
we drop a common factor $(4\pi E R^3(t)/R_0^3)^{-1}$ on both sides
of the kinetic equations.}

\begin{equation}
\label{L[n]} {\cal
L}[n_{\nu_i}(E,t)]=\left[\frac{\partial}{\partial t}-H(t)E
\frac{\partial}{\partial E}-H(t)\right]n_{\nu_i}(E,t)\ .
\end{equation}

The collision operator of unstable neutrinos reads
\begin{equation}
 {\cal C}[n_{\nu_i}(E,t)] =
R_{\mathrm{SN}}(t)Y_{\nu_i}(E)  + \sum_{m_j>m_i} q_{ji}(E,t) -
\Gamma_i \,\frac{m_i}{E}\,n_{\nu_i}(E,t) \label{Coll}\ ,
\end{equation}
where
\begin{equation}
\label{qji} q_{ji}(E,t)=\int_E^\infty dE'
n_{\nu_j}(E',t)\,\Gamma(\nu_j\to\nu_i)\,\frac{m_j}{E'}\,\psi_{\nu_j\to\nu_i}(E',E)\
.
\end{equation}
The right-hand side (r.h.s.) of Eq.~(\ref{Coll}) contains  two
{\em source\/} terms and one {\em sink\/} term. The first source
term quantifies the standard (decay-independent) emission of
$\nu_i$ from core-collapse supernovae. The second source term
quantifies the population increase of $\nu_i$ due to decays from
heavier states; this term is absent for the heaviest neutrino
state(s). The last (sink) term on the right-hand side (r.h.s.) of
Eq.~(\ref{Coll}) represents the loss of $\nu_i$ due to decay to
lighter states with total width $\Gamma_i$; this term is absent
for the lightest, stable neutrino state(s).%
\footnote{Additional terms in the collision operator, due to Pauli
blocking and to inverse reactions $X+\nu_j\to\nu_i$, can be safely
neglected at the very low number densities of SRN.}

\subsection{General solution}

In order to solve the set of kinetic equations
(\ref{Kinetic})--(\ref{qji}), we first rewrite them in terms of
the redshift variable $z=z(t)$ and of the redshifted energy
$\varepsilon=\varepsilon(E,z)$,
\begin{equation}
\label{Change} \left(\begin{array}{c}t\\E\end{array}\right)\to
\left(\begin{array}{c}z\\ \varepsilon\end{array}\right)\equiv
\left(\begin{array}{c} R_0/R(t)-1\\E/(1+z)\end{array}\right) \ ,
\end{equation}
with associated partial derivatives
\begin{equation}
\label{Partials}
\left(\begin{array}{c}
\frac{\partial}{\partial t}\\
\\
\frac{\partial}{\partial E}
\end{array}\right)=
\left(\begin{array}{cc}
\frac{\partial z}{\partial t} & \frac{\partial \varepsilon}{\partial t}\\
\\
\frac{\partial z}{\partial E} & \frac{\partial \varepsilon}{\partial E}\\
\end{array}\right)
\left(\begin{array}{c}
\frac{\partial}{\partial z}\\
\\
\frac{\partial}{\partial\varepsilon}
\end{array}\right)=
\left(\begin{array}{c} -H(z)\left(
(1+z)\frac{\partial}{\partial z}-\varepsilon\frac{\partial}{\partial\varepsilon}\right)\\
\\
\frac{1}{1+z}\frac{\partial}{\partial\varepsilon}
\end{array}\right)\ ,
\label{eq:derivates}
\end{equation}
where the relation $\dot{z}=-(1+z)H(z)$ has been used. With this
change of variables, the Liouville operator depends on $z$ only,
and the kinetic equations can be directly integrated.

More precisely, by defining the auxiliary function
\begin{equation}
\label{csi} \xi(z)=\int_0^zdz'\,H^{-1}(z')(1+z')^{-2}\ ,
\end{equation}
and the global source term
\begin{equation}
\label{Source}
S_i(\varepsilon(1+z),z)=R_\mathrm{SN}(z)Y_{\nu_i}(\varepsilon(1+z))
+\sum_{m_j>m_i}q_{ji}(\varepsilon(1+z),z)\ ,
\end{equation}
the neutrino kinetic equations (\ref{Kinetic})--(\ref{Coll}) can
be cast in a compact form
\begin{equation}
-H(z)\,e^{m_i\Gamma_i\xi(z)/\varepsilon}\,\frac{\partial}{\partial
z}\left[(1+z)n_{\nu_{i}}\,e^{-m_i\Gamma_i\xi(z)/\varepsilon}
\right] = S_i(\varepsilon(1+z),z)\label{compact}\ ,
\end{equation}
which is easily integrated, giving
\begin{equation}
\label{Niint} n_{\nu_i}=\frac{1}{1+z} \int_z^\infty
\frac{dz'}{H(z')}\,S_i(\varepsilon(1+z'),z')\,
e^{-m_i\Gamma_i[\xi(z')-\xi(z)]/\varepsilon}\ .
\end{equation}
By replacing back the variable $E=\varepsilon(1+z)$, one obtains
the general solution of the neutrino kinetic equations,
\begin{eqnarray}
n_{\nu_i}(E,z) &=& \frac{1}{1+z}\int_{z}^{\infty}
\frac{dz'}{H(z')}\,\left[
R_\mathrm{SN}(z')\,Y_{\nu_i}\left(E\frac{1+z'}{1+z}\right) \right.\nonumber\\
& & +\left.\sum_{m_j>m_i}q_{ji}\left(E\frac{1+z'}{1+z},z'\right)
\right]e^{-m_i\Gamma_i[\xi(z')-\xi(z)](1+z)/E}\ , \label{Solution}
\end{eqnarray}
which holds for generic neutrino mass spectra $m_i$, decay widths
$\Gamma_i$, and decay energy spectra $\psi_{\nu_j\to\nu_i}(E,E')$.
The effect of flavor transitions is embedded in the yields
$Y_{\nu_i}$ (see Table~I), in the same way as for the no-decay
case discussed in Sec.~II. Notice that the dependence of $1/H(z')$
upon the cosmology cancels with the factor $R_\mathrm{SN}(z')$
(see Sec.~II~A) but not with the other factors in
Eq.~(\ref{Solution}). Therefore, in the presence of decay, the SRN
density acquires a dependence on the cosmological parameters
$(\Omega_M,\Omega_\Lambda)$.

The double integration (over energy and redshift) implied by
Eqs.~(\ref{qji})  and (\ref{Solution}) can be performed
numerically by following the decay sequence, i.e., starting from
the heaviest state ($q_{ji}=0$) and ending at the lightest state
($\Gamma_i=0$). The observable local supernova relic $\nu_i$
density is finally obtained by setting $z=0$ in $n_{\nu_i}(E,z)$.
We conclude by noting that the case of no decay [Eq.~(\ref{ni})]
is obtained from the general solution [Eq.~(\ref{Solution})] at
$z=0$ in the limit $\tau_i\to\infty$ (i.e., $\Gamma_i=0=q_{ji}$),
as expected.

\section{Applications to scenarios inspired by Majoron models}

In this section we apply the general results of Sec.~III~B to some
representative decay scenarios, inspired by Majoron models. After
a brief overview of the $2\nu$ decay case, we examine and compare
a few representative $3\nu$ decay cases, which provide SRN yields
higher, comparable, or lower than for no decay. Simplificative
assumptions will be made in all cases, in order to reduce the
parameter space, and to highlight the main effects of neutrino
decay.

\subsection{Two-family decay}

Let us consider a doublet of ``heavy'' and ``light''  neutrino
mass eigenstates $\nu_{h,l}$ ($m_h>m_l$).  We briefly recall that
nonradiative (invisible) decays of the kind
\begin{equation}
\label{DecayHL}\nu_h\to\nuornubar_l+X\ ,
\end{equation}
may arise through the coupling of $\nu_{h,l}$ to a very light
scalar or pseudoscalar particle $X$ (assumed to be effectively
massless for our purposes). In particular, $X$ can be the
Goldstone boson (Majoron) in models with spontaneous violation of
the $B-L$ symmetry of the standard electroweak model \cite{Pecc}.

There is a vast literature on neutrino-Majoron decay models (see,
e.g., \cite{Gelm,Dolg} and references therein) and related
phenomenological constraints on neutrino-Majoron couplings (see,
e.g., \cite{Pakv,PaVa,Kach,Toma,Raft,Farz}). Since the
neutrino-Majoron coupling can also contribute to the neutrino
mass, model-dependent relations may arise between neutrino mass
and decay parameters (see, e.g., \cite{PaVa,Toma}). The branching
ratios and final-state spectra for the two channels
$\nu_h\to\nu_l$ and $\nu_h\to\bar\nu_l$ are also functions of
model-dependent couplings.

Here, however, we do not commit ourselves to any specific
theoretical model, and assume that the lifetime-to-mass ratio
$\tau/m_h$ is a free parameter [subject only to the safe
constraint in Eq.~(\ref{solarbound}), when needed]. We also focus
on two phenomenologically interesting cases in which the branching
ratios and decay spectra become model-independent, namely, the
case of quasidegenerate (QD) neutrino masses ($m_h\simeq m_l\gg
m_h-m_l$) and of strongly hierarchical (SH) neutrino masses ($m_h
\gg m_l\simeq 0$). Table~II displays the relevant characteristics
of the QD and SH cases (for ultrarelativistic neutrinos), obtained
as appropriate limits of the general results worked out in
Ref.~\cite{KLam}.

\begin{table}[t]
\caption{\footnotesize\baselineskip=4mm Branching ratios and
energy spectra in $2\nu$  Majoron decay scenarios
$\nu_h\to\nuornubar_l+X$, for the extreme cases of
quasi-degenerate (QD) and strongly hierarchal (SH) masses of the
``heavy'' and ``light'' active neutrinos $\nu_{h,l}$. Analogous
expressions hold for $\bar\nu_h$ decay, with the replacements
$\nu_k\leftrightarrow \bar\nu_k$ ($k=h,l$).
\smallskip}
\begin{ruledtabular}
\begin{tabular}{lc|cc|cc}
Case & Mass relations & $B(\nu_h\to\nu_l)$ &
$B(\nu_h\to\bar\nu_l)$~~~ & $\psi_{\nu_h\to\nu_l}(E_h,E_l)$ &
$\psi_{\nu_h\to\bar\nu_l}(E_h,E_l)$~~ \\[1mm]
\hline
 QD & \phantom{\Bigg|} $m_h\simeq m_l\gg m_h-m_l$~~ & 1 & 0 &
$\delta (E_h-E_l)$ & ---
\\[2mm]
 SH & $m_h\gg m_l\simeq 0$ & $1/2$ & $1/2$ &
$\displaystyle\frac{2 E_l}{E_h^2}$
& $\displaystyle\frac{2}{E_h}\left(1-\frac{E_l}{E_h}\right)$\\
\end{tabular}
\end{ruledtabular}
\end{table}

In the QD case (Table~II), the $\nu_h$ decays only into $\nu_l$,
which carries the whole $\nu_h$ energy. When the decay is {\em
complete\/} (i.e., for lifetimes $\tau E/m_h\ll H_0^{-1}$ in the
laboratory frame), the initial $\nu_h$ energy spectrum is then
directly transferred to the $\nu_l$ energy spectrum.

In the SH case, the $\nu_h$ decays into either $\nu_l$ or
$\bar\nu_l$ with the same probability, but with  different energy
distributions (Table~II). When the decay is complete, the initial
$\nu_h$ energy spectrum is then transferred to the final $\nu_l$
and $\bar\nu_l$ spectra through convolutions with the
$\psi$-functions in Table~II. For later purposes, in the SH case
we define two (normalization-preserving) convolution operators
$\mathcal{D}$ and $\overline\mathcal{D}$, acting over the $\nu_h$
yield functions $Y_{\nu_h}(E)$ as:
\begin{eqnarray}
{\cal D}Y_{\nu_h}(E) &=& \int_E^\infty
dE'\,\psi_{\nu_h\to\nu_l}(E',E)Y_{\nu_h}(E')\ ,\label{calD}\\
\overline{\cal D}Y_{\nu_h}(E) &=& \int_E^\infty
dE'\,\psi_{\nu_h\to\bar\nu_l}(E',E)Y_{\nu_h}(E')\ .\label{calDbar}
\end{eqnarray}
Notice that, for supernova neutrino yields parameterized through
Eq.~(\ref{varphi}) \cite{Spec}, the specific choice $\alpha=3$ in
Eq.~(\ref{alpha}) makes the integrals in Eqs.~(\ref{calD}) and
(\ref{calDbar}) analytical and
elementary.%
\footnote{ For such choice, operators of the kind ${\cal
D}\overline{\cal D}+ \overline{\cal D}{\cal D}$ and ${\cal
D}^2+\overline{\cal D}{}^2$, which will appear in $3\nu$ decay
cases, also lead to analytical integrals (in terms of exponential
integral functions). For generic, noninteger values of $\alpha$,
the  analytical expressions become rather involved, and numerical
evaluations are preferable.}
Qualitatively, in the SH case the action of the operators $\cal D$
and $\overline{\cal D}$ is to increase the neutrino yields at low
energy.

The above notation, although introduced for the $2\nu$ decay case,
will be frequently used in the following subsections. In fact, we
shall consider  $3\nu$ scenarios whose $2\nu$ sub-decays can be
treated within either the QD or the SH approximation.

\subsection{Three-family decays for normal hierarchy and quasidegenerate
masses}

In this section we consider a representative decay scenario which
provides SRN densities generally {\em higher\/} than for no decay.
The scenario involves normal mass hierarchy ($+\Delta m^2$), with
masses much larger than their splittings ($m^2_i\gg \Delta
m^2,\,\delta m^2$). The quasidegenerate (QD) approximation, which
forbids decays of neutrinos into antineutrinos and vice versa, can
thus be applied (see Table~II). In the context of supernova relic
$\bar\nu_e$'s, we shall consider only antineutrino decays
($\bar\nu_3\to \bar\nu_{1,2}$ and $\bar\nu_2\to\bar\nu_1$).

For the sake of simplicity, we assume that
$B(\bar\nu_3\to\bar\nu_2)=B(\bar\nu_3\to\bar\nu_1)=1/2$, and also
that $\tau_3/m_3=\tau_2/m_2\equiv\tau/m$. By construction, the
decay scenario considered in this section is thus governed by just
one
free parameter%
\footnote{Oscillation parameters have been fixed previously, and
the only relevant unknown ($\theta_{13}$ or, equivalently, $P_H$)
does not affect antineutrinos in normal hierarchy (see Table~I).}
($\tau/m$). Notice that, for $\tau/m\sim O(10^{10})$~s/eV, SRN
decay effects are expected to occur on a truly cosmological scale
[see Eq.~(\ref{upperbound})]. For much larger values of $\tau/m$,
the no-decay case is recovered. For much smaller values of
$\tau/m$, SRN decay is instead {\em complete\/},  all SRN being in
the lightest mass eigenstate $\bar\nu_i$ at the time of detection.

Figure~3 shows the supernova relic $\bar\nu_e$ energy spectrum,
and the associated (observable) positron spectrum, for the decay
scenario described above (and graphically shown in the left
panel). The energy spectra for complete decay (red solid curves)
appear to be a factor of $\sim 2$ higher---and also slightly
harder---than for no decay (black solid curve). This difference is
entirely due to the role of $\nu_x$, as explained below.

\begin{figure}
\vspace*{-0.0cm}\hspace*{-0.2cm}
\includegraphics[scale=0.95]{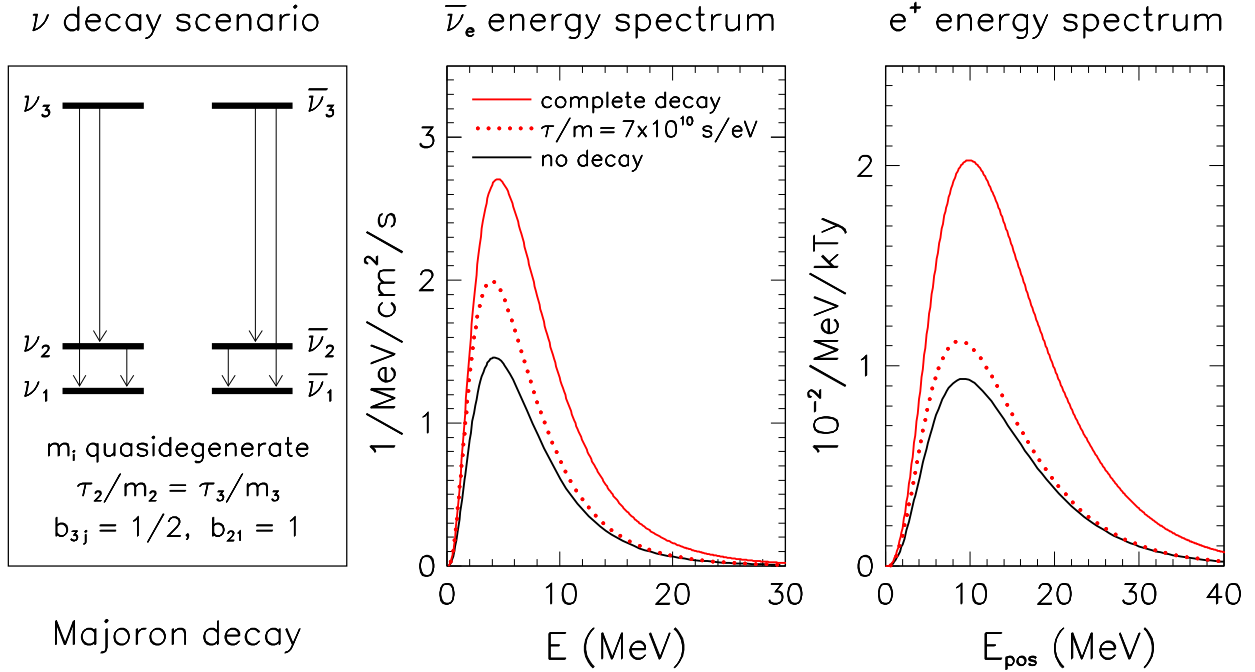}
\vspace*{-0.2cm} \caption{\label{fig3}
\footnotesize\baselineskip=4mm (Color online.) Supernova relic
$\bar\nu_e$ spectrum (middle panel), and associated positron
spectrum from $\bar\nu_e+p\to n+e^+$ (right panel), for a decay
scenario with normal hierarchy and quasidegenerate masses (left
panel, with $\tau/m$ and branching ratios assignments). The red
and black solid curves correspond to the limiting cases of
complete decay and no decay, respectively. The red dotted curves
correspond to incomplete decay with $\tau/m=7\times 10^{10}$~s/eV.
See the text for details.}
\end{figure}

For complete decay, the final state is populated by stable
$\bar\nu_1$'s only, coming both from the initial $\bar\nu_1$
component ($Y_{\bar\nu_1}=Y_{\bar\nu_e}$) and from fully decayed
$\bar\nu_{2,3}$'s ($Y_{\bar\nu_2}+Y_{\bar\nu_3}=2Y_{\nu_x}$, see
Table~I), with unaltered neutrino energies (QD case in Table~II).%
\footnote{The case of complete decay could also be obtained from
the general solution in the limit $\tau/m\to 0$ (derivation
omitted).}
 The final relic $\bar\nu_e$ density (given by
$\cos^2\theta_{12}$ times the final density of $\bar\nu_1$) is
thus obtained by redshifting an initial yield equal to
$\cos^2\theta_{12}Y_{\bar\nu_e} +2 \cos^2\theta_{12} Y_{\nu_x}$.
In the case of no decay for normal hierarchy, the initial yield is
instead given by $\cos^2\theta_{12}\,Y_{\bar\nu_e}
+\sin^2\theta_{12}Y_{\nu_x}$ [see Eq.~(\ref{ni2})]. Therefore,
while the $Y_{\bar\nu_e}$ component is the same in the two cases,
the weight of the $Y_{\nu_x}$ component for complete decay is
$2\cos^2\theta_{12}\simeq 1.42$, much larger than for no decay,
where the weight is $\sin^2\theta_{12}\simeq 0.29$. The stronger
weight of $Y_{\nu_x}$ for complete decay leads to the increase in
normalization, peak energy, and width, which is
visible in Fig.~3.%
\footnote{The enhancement of the SRN yield for the decay scenario
with normal hierarchy and quasidegenerate masses has also been
discussed in Ref.~\cite{Ando}.}

For incomplete neutrino decay (i.e., for $\tau/m\sim
O(10^{10})$~s/eV), one expects an intermediate situation leading
to a SRN flux moderately higher than for no decay. Figure~3
displays the results for a representative case ($\tau/m=7\times
10^{10}$~s/eV, red dotted curves), as obtained through the general
solution of the kinetic equations worked out in Sec.~III.
Summarizing, the decay scenario examined in this section can lead
to an increase of the SRN rate, as compared with the case of no
decay. The enhancement can be as large as a factor $\sim 2$, the
larger the more complete is the decay.

\subsection{Three-family decays for normal hierarchy and $m_1\simeq 0$}

In this section we consider a representative decay scenario which
provides observable SRN densities generally {\em comparable\/} to
the no-decay case. The scenario assumes that the mass hierarchy is
normal ($+\Delta m^2$) and that the lightest state is basically
massless ($m_1\simeq 0$), so that $m_{2,3}\gg m_1$ and the
approximation of strong hierarchy (SH) can be applied to the
decays of $\nu_{2,3}$ (and of $\bar\nu_{2,3}$).

According to Table~II (SH case), we can take
$B(\nu_2\to\nu_1)=B(\nu_2\to\bar\nu_1)=1/2$. For the sake of
simplicity, we extend such ``branching ratio democracy''  to all
the $\nu_3$ decay channels, namely, we take
$B(\nu_3\to\nu_1)=B(\nu_3\to\bar\nu_1)=B(\nu_3\to\nu_2)=
B(\nu_3\to\bar\nu_2)=1/4$. We also assume
$\tau_3/m_3=\tau_2/m_2\equiv\tau/m$, so that, as in the previous
section, there is only one free parameter ($\tau/m$). The cases of
no decay, incomplete decay, and complete decay, are then obtained
for $\tau/m$ much larger, comparable, or much smaller than
$O(10^{10})$~s/eV, respectively.

Figure~4 shows the supernova relic $\bar\nu_e$ and positron
spectra for this scenario where, as depicted in the left panel,
all decay channels are open. The results for the no-decay limit
(solid curves) are identical to those in Fig.~3 and are not
discussed again. The middle panel in Fig.~4 shows that the
neutrino spectrum for complete decay (green solid curve) is
significantly enhanced at low energy, as compared with the one for
no decay. The positron spectrum, however, is very similar in the
two cases (right panel). These features can be understood as
follows.

\begin{figure}[t]
\vspace*{-0.0cm}\hspace*{-0.2cm}
\includegraphics[scale=0.95]{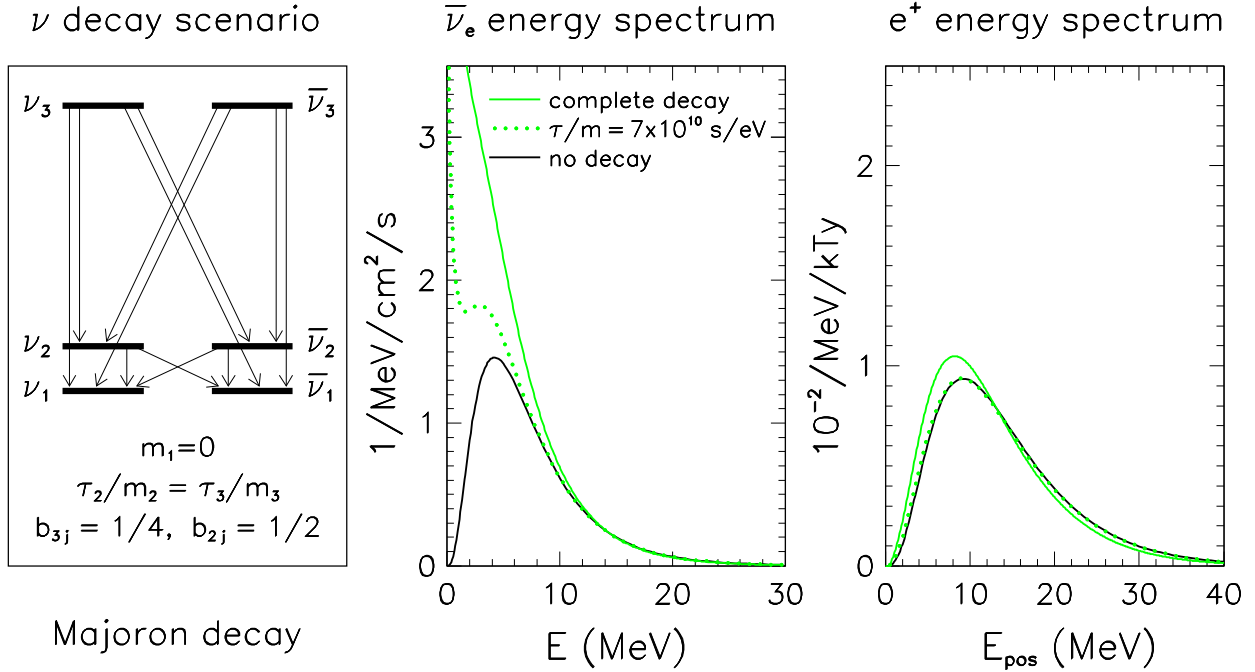}
\vspace*{-0.7cm} \caption{\label{fig4}
\footnotesize\baselineskip=4mm (Color online.) Supernova relic
$\bar\nu_e$ spectrum (middle panel), and associated positron
spectrum from $\bar\nu_e+p\to n+e^+$ (right panel), for a decay
scenario with normal hierarchy and $m_1\simeq 0$ (left panel, with
$\tau/m$ and branching ratios assignments). The green and black
solid curves correspond to the limiting cases of complete decay
and no decay, respectively. The green dotted curves correspond to
incomplete decay with $\tau/m=7\times 10^{10}$~s/eV. See the text
for details.}
\end{figure}

For complete decay, the final yield of stable $\bar\nu_1$'s comes
both from the initial $\bar\nu_1$'s and from a complex decay
chain, through the action of the operators $\cal D$ and
$\overline{\cal D}$ [Eqs.~(\ref{calD}) and (\ref{calDbar})]:
\begin{eqnarray}
\bar\nu_1 \longrightarrow\bar\nu_1 &\Longrightarrow& Y_{\bar\nu_1}\label{first}\ ,\\
\bar\nu_2\stackrel{\frac{1}{2}\cal D}{\longrightarrow}\bar\nu_1
&\Longrightarrow& \frac{1}{2}{\cal D}Y_{\bar\nu_2}\ ,\\
\nu_2\stackrel{\frac{1}{2}\overline{\cal
D}}{\longrightarrow}\bar\nu_1
&\Longrightarrow& \frac{1}{2}{\overline{\cal D}}Y_{\nu_2}\ ,\\
\bar\nu_3\stackrel{\frac{1}{4}\cal D}{\longrightarrow}\bar\nu_1
&\Longrightarrow& \frac{1}{4}{\cal D}Y_{\bar\nu_3}\ ,\\
\nu_3\stackrel{\frac{1}{4}\overline{\cal
D}}{\longrightarrow}\bar\nu_1
&\Longrightarrow& \frac{1}{4}{\overline{\cal D}}Y_{\nu_3}\ ,\\
\bar\nu_3 \stackrel{\frac{1}{4}\cal
D}{\longrightarrow}\bar\nu_2\stackrel{\frac{1}{2}\cal
D}{\longrightarrow}\bar\nu_1
&\Longrightarrow& \frac{1}{8}{\cal D}\,{\cal D}Y_{\bar\nu_3}\ ,\\
\bar\nu_3 \stackrel{\frac{1}{4}\overline{\cal
D}}{\longrightarrow}\nu_2\stackrel{\frac{1}{2}\overline{\cal
D}}{\longrightarrow}\bar\nu_1
&\Longrightarrow& \frac{1}{8}{\overline{\cal D}}\,{\overline{\cal D}}Y_{\bar\nu_3}\ ,\\
\nu_3 \stackrel{\frac{1}{4}\overline{\cal
D}}{\longrightarrow}\bar\nu_2\stackrel{\frac{1}{2}\cal
D}{\longrightarrow}\bar\nu_1
&\Longrightarrow& \frac{1}{8}{\cal D}\,{\overline{\cal D}}Y_{\nu_3}\ ,\\
\nu_3 \stackrel{\frac{1}{4}\cal
D}{\longrightarrow}\nu_2\stackrel{\frac{1}{2}\overline{\cal
D}}{\longrightarrow}\bar\nu_1 &\Longrightarrow&
\frac{1}{8}{\overline{\cal D}}\,{\cal D}Y_{\nu_3}\label{last}\ .
\end{eqnarray}
This chain leads to a SRN density equation for complete decay
formally similar to Eq.~(\ref{ni2}) (no decay), but with the
integrand in square brackets replaced by $\cos^2\theta$ times the
sum of the terms in Eqs.~(\ref{first})--(\ref{last}):
\begin{equation}
\cos^2\theta_{12}\left( Y_{\bar\nu_1}+ \frac{1}{2}{\cal D
}Y_{\bar\nu_2}+\dots+\frac{1}{8}{\overline{\cal D}}{\cal
D}Y_{\nu_3}\right)\ .
\end{equation}

In the above expression for complete decay, terms which are linear
and quadratic in the convolution operators produce a substantial
enhancement of the SRN energy spectrum at low energy, visible as a
``pile-up'' of decayed neutrinos with degraded
energy in the middle panel of Fig.~4.%
\footnote{The terms linear and quadratic in $\cal D$ and
$\overline{\cal D}$ carry a mild dependence on the crossing
probability $P_H$ through the neutrino yields $Y_{\nu_i}$'s (see
Table~I for normal hierarchy). Figure~4 refers to the case
$P_H=0$; very similar results are obtained for $P_H=1$ (not
shown).}
 At high energy, however, such terms happen to provide a
contribution which is numerically close to the no-decay term
$\sin^2\theta_{12}Y_{\nu_x}$ in Eq.~(\ref{ni2}), so that the
high-energy tails of the SRN spectra for no decay and for complete
decay (black and green solid curves in the middle panel of Fig.~4)
are accidentally very similar to each other. Analogously, the case
of incomplete decay (e.g., $\tau/m=7\times 10^{10}$~s/eV, green
dotted curves), is appreciably different from the cases of no
decay and of complete decay only at low energy.

In this scenario, the interesting effects of decay are almost
completely confined to low $\bar\nu_e$ energies, and are thus
washed out in the observable $e^+$ spectrum, due to the cross
section enhancement of high-energy features. In fact, the $e^+$
spectra for the three cases of complete, incomplete, and no decay,
turn out to be very similar to each other (right panel of Fig.~4).

In conclusion, from the results of this section we learn that
there are neutrino decay scenarios which, despite a relatively
complex structure (see Fig.~4), cannot be distinguished from the
no decay case through future SRN observations, for any value of
$\tau/m$ above the safe bound in Eq.~(\ref{solarbound}).
Similarly, such scenarios do not alter the ``standard'' (no-decay)
SN 1987A phenomenology, and thus provide a particularly clean
example of how the naive limit in Eq.~(\ref{SN87bound}) can
actually be evaded.


\begin{figure}[t]
\vspace*{-0.0cm}\hspace*{-0.2cm}
\includegraphics[scale=0.95]{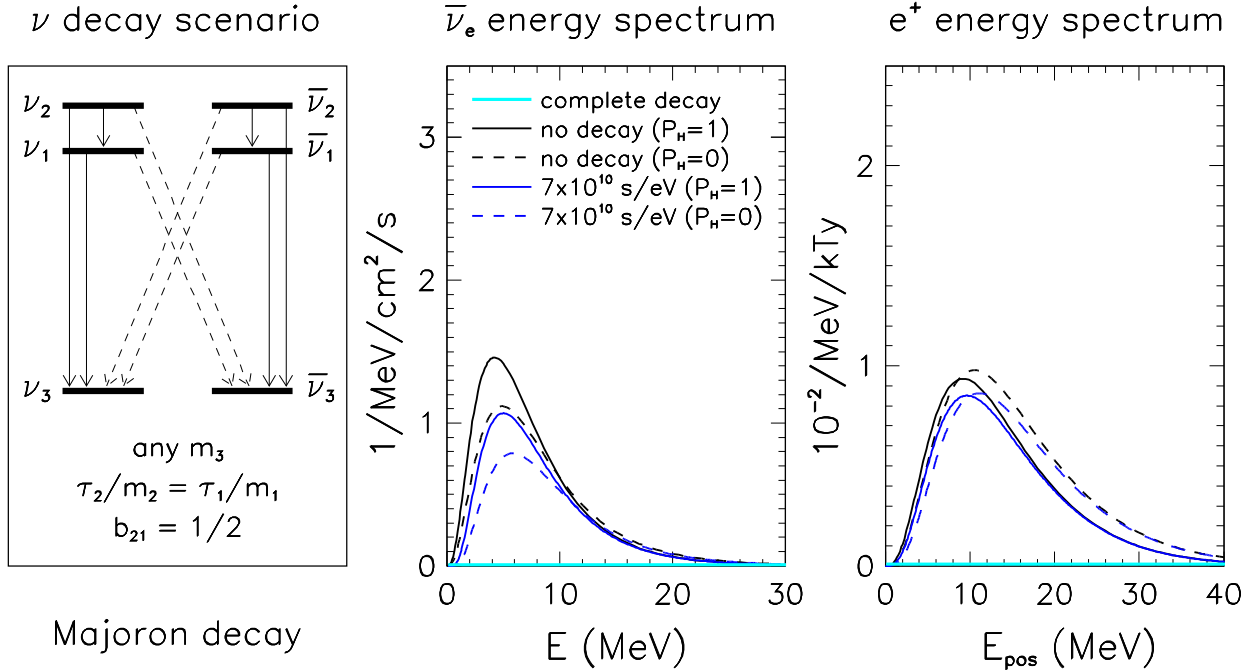}
\vspace*{-0.7cm} \caption{\label{fig5}
\footnotesize\baselineskip=4mm (Color online.) Supernova relic
$\bar\nu_e$ spectrum (middle panel), and associated positron
spectrum from $\bar\nu_e+p\to n+e^+$ (right panel), for a decay
scenario with inverted hierarchy and generic $m_3$ (left panel,
with $\tau/m$ and branching ratios assignments). The black and the
blue curves correspond, respectively, to no decay and to
incomplete decay with $\tau/m=7\times 10^{10}$~s/eV, for both
$P_H=1$ (solid) and $P_H=0$ (dashed). The horizontal (light blue)
line at $\sim 0$ corresponds to the case of complete decay. See
the text for details.}
\end{figure}

\subsection{Three-family decay for inverted hierarchy}

We conclude our survey of $3\nu$ decays  by discussing a scenario
where the SRN density is generally {\em suppressed\/}, as compared
with the case of no decay. The scenario assumes an inverted mass
hierarchy ($-\Delta m^2$), a fixed branching ratio
$B(\bar\nu_2\to\bar\nu_1)=1/2\ [=B(\nu_2\to\nu_1)]$, and
$\tau_1/m=\tau_2/m\equiv\tau/m$. Under these assumptions, only the
decay $\bar\nu_2\to\bar\nu_1$ (where the QD approximation is
applicable) is relevant to SRN observations. In fact, decays to
$\bar\nu_3$ provide a negligible amount of $\bar\nu_e$'s ($\propto
\sin^2\theta_{13}$). In particular, as graphically reported in the
left panel of Fig.~5,  the absolute value of $m_3$ makes no
difference: It only opens (closes) the ``dashed'' decay channels
for $m_3\simeq 0$ ($m_3$ large), with no change on the relic
$\bar\nu_e$ flux for fixed $B(\bar\nu_2\to\bar\nu_1)$. In the
above scenario, the case of complete decay ($\tau/m\ll
O(10^{10})$~s/eV) is trivial: since the final state is populated
only by $\bar\nu_3$ (and $\nu_3$), the relic density of
$\bar\nu_e$
is negligibly small ($\propto \sin^2\theta_{13}$).%
\footnote{This is true, in general, for complete decay in any IH
scenario.}

The nontrivial case of incomplete decay ($\tau/m\sim
O(10^{10})$~s/eV) is then expected to lead to an intermediate
suppression of the SRN density. Figure~5 shows the numerical
results for the specific value $\tau/m=7\times 10^{10}$~s/eV, in
both cases $P_H=1$ (solid curves) and $P_H=0$ (dashed curves). The
neutrino spectra for incomplete decay (blue curves) appear to be
systematically lower than the corresponding no-decay spectra
(black curves), although the difference is mitigated in the
positron spectra (right panel of Fig.~5). A more substantial
suppression of the positron spectrum can be obtained by lowering
$\tau/m$ (see next subsection). In conclusion, in the inverted
hierarchy scenario of Fig.~5, the SRN signal is generally
suppressed by neutrino decay, and can eventually disappear for
complete decay.

\subsection{Overview and summary of $3\nu$ decay}

In the previous three subsections we have discussed, for three
rather different scenarios, the cases of no decay [$(\tau/m\gg
O(10^{10})$~s/eV], of complete decay [$\tau/m \ll
O(10^{10})$~s/eV], and of incomplete decay for a specific value of
$\tau/m$ ($=7\times 10^{10}$~s/eV). We think it useful to show
also the behavior of the SRN signal for continuous values of the
free parameter $\tau/m$.%
\footnote{This task implies extensive calculations through the
general solution worked out in Sec.~III.}

Figure~6 shows the positron event rate integrated in the energy
window $E_{e^+}\in [10,20]$ MeV which, although lower than the
current SK window \cite{SKSN}, might become accessible to future,
low-background SRN searches \cite{GADZ}. For each scenario, the
rate is normalized to the standard expectations for no decay and
normal hierarchy (NH), and is plotted as a function of $\tau/m$.
The black solid line represents the reference case (no decay in
NH), which is indistinguishable from the case of no decay with
inverted hierarchy  (IH) and $P_H=1$ (see Sec.~II~D). The black
dashed line refers instead to the no-decay case with IH and
$P_H=0$. For no decay, variations in the hierarchy or in $P_H$
appear  to induce relatively small effect, which might be
difficult to uncover once realistic experimental and theoretical
uncertainties are considered in future SRN observations. Decay
effects, however, enlarge dramatically the possible range of
observable positron rates provided that $\tau/m$ is in the
cosmologically interesting range below $O(10^{11})$~s/eV [See
Eq.~(\ref{upperbound})]. For $\bar\nu$ decay in the
quasidegenerate NH spectrum of Sec.~IV~B (red curve in Fig.~6),
the positron rate rapidly increases for decreasing $\tau/m$, and
reaches the ``complete decay plateau'' already at $\tau/m\sim
O(10^9)$~s/eV, with an asymptotic enhancement by a factor $\sim
2.3$. In the case of IH spectrum considered in Sec.~IV~D (solid
and dashed blue curves in Fig.~6), conversely, the positron rate
vanishes when approaching $\tau/m\sim O(10^9)$~s/eV, for both
cases $P_H=1$ (solid) and $P_H=0$ (dashed). Finally, for the NH
spectrum with $m_1\simeq 0$ of Sec.~IV~C (green curve in Fig.~6),
the positron rate appears to be almost the same as for no decay,
at any value of $\tau/m$.

Summarizing, neutrino decay can enlarge the reference no-decay
predictions for observable positron rates by any factor $f$ in the
range $\sim [0,2.3]$.%
\footnote{These results refer to a prospective positron energy
window $E_{e^+}\in[10,20]$~MeV. For the current SK analysis window
$E_{e^+}\in[18,34]$~MeV, the range would be very similar, $f\in
[0,2.7]$. Variations of the reference inputs in Sec.~II can also
lead to wider ranges.}

Since the current experimental upper bound on the SRN flux from SK
\cite{SKSN} is just a factor of $\sim 2$--3 above typical no-decay
expectations \cite{SKSN,An03} (including those in this work),
future observations below such bound are likely to have an impact
on neutrino decay models. If experimental and theoretical
uncertainties can be kept smaller than a factor of two (a
nontrivial task), one should eventually be able to rule out, at
least, either the lowermost or the uppermost values in the range
$f\in [0,2.3]$, i.e., one of the extreme cases of ``complete
decay.'' Optimistically, one might then try to constrain specific
decay models and lifetime-to-mass ratios through observations (a
goal not reachable with current information \cite{Ando}).
Degeneracy between decay and no decay in specific models (as the
one considered in Sec.~IV~C) will, however, set intrinsic
limitations to SRN tests of neutrino lifetimes.

\begin{figure}[t]
\vspace*{-0.0cm}\hspace*{-0.2cm}
\includegraphics[scale=0.58]{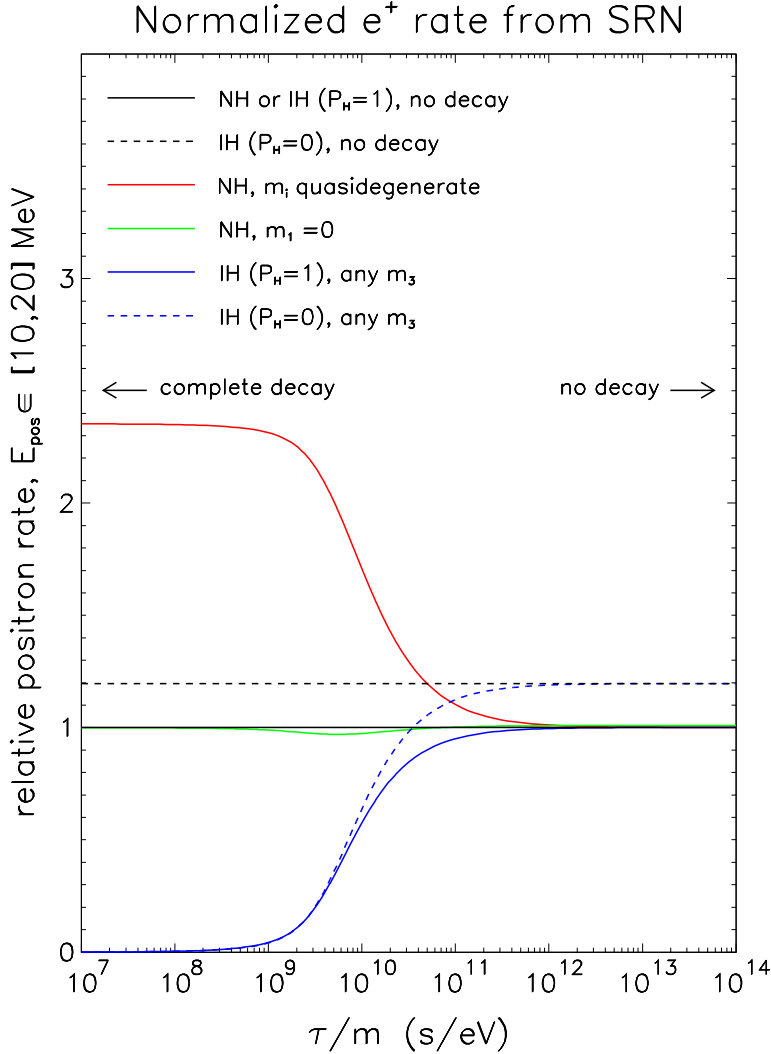}
\vspace*{+0.0cm} \caption{\label{fig6}
\footnotesize\baselineskip=4mm (Color online.) Positron event
rates in the energy range $[10,20]$~MeV for various decay
scenarios, normalized to standard expectations for normal
hierarchy and no decay. The $\tau/m$ range in abscissa is well
above the safe bound in Eq.~(\ref{solarbound}). Notice how the
expectations branch out (and then reach the complete decay limit)
in the cosmologically relevant range $\tau/m\lesssim
10^{11}$~s/eV. See the text for details. }
\end{figure}

\section{Conclusions and PROSPECTS}

Neutrino decays with cosmologically relevant neutrino lifetimes
[$\tau_i/m_i\lesssim O(10^{11})$~s/eV] can, in principle, be
probed through observations of supernova relic $\bar\nu_e$ (SRN).
We have shown how to incorporate the effects of both flavor
transitions and decays in the calculation of the SRN density, by
finding the general solution of the neutrino kinetic equations for
generic two-body nonradiative decays. (Radiative decays are
briefly commented upon in Appendix~A.) We have then applied such
solution to three representative decay scenarios which lead to an
observable SRN density larger, comparable, or smaller than for no
decay. In the presence of decay, the expected range of the SRN
rate is significantly enlarged (from zero up to the current upper
bound). Future SRN observations can thus be expected to constrain
at least some extreme decay scenarios and, in general, to test the
likelihood of specific decay models, as compared with the no-decay
case.

In this work we have focused on theoretical SRN calculations, and
have not attempted a phenomenological analysis of the sensitivity
to neutrino decay through {\em prospective\/} SRN experimental
data, as those which might be collected, e.g., in the proposed
Gd-doped Super-Kamiokande detector \cite{GADZ}, in
Hyper-Kamiokande \cite{Hype}, or in the UNO project \cite{UNNO}.
Such analysis would require a detailed characterization of the
many uncertainties affecting both the background-subtracted SRN
signal and the theoretical inputs, especially those related to
supernova simulations and to the star formation rate. These
uncertainties, although currently rather large, are likely to
decrease in the future as more powerful supernova codes and new
astrophysical observations will become available (including,
hopefully, a galactic supernova explosion). When significant
improvements will be made in this direction, a systematic approach
to SRN predictions and to their uncertainties will certainly be
required, in order to interpret future SRN data and to use them to
constrain nonradiative neutrino lifetimes in a range [up to
$\tau/m\sim O(10^{11})$~s/eV] which is both largely unexplored and
difficult to access by other means.

\acknowledgments We thank M.\ Malek for helpful correspondence
about supernova relic neutrinos in Super-Kamiokande. G.L.F.\ and
D.M.\ acknowledge useful discussions on SRN with several
participants to the Workshop ``Thinking, Observing and Mining the
Universe'' (Sorrento, Italy, 2003). This work is supported in part
by the Istituto Nazionale di Fisica Nucleare (INFN) and by the
Italian Ministry of Education (MIUR) through the ``Astroparticle
Physics'' project.

\appendix
\section{Radiative decays of supernova relic neutrinos}

Radiative neutrino decays \cite{El74,Pe77} of the kind
\begin{equation}
\label{radiat} \nuornubar_j \to\ \nuornubar_i + \gamma\
\end{equation}
have been considered in a number of papers (see, e.g.,
\cite{Dolg,Pakv,PaVa,Raft} and references therein). Limits on the
neutrino lifetime/mass ratio $\tau/m$  for such decay modes can be
derived from a variety of arguments \cite{Raft}. E.g., one of the
strongest bounds,
\begin{equation}
\label{stronglimit} \tau/m\gtrsim O(10^{20})\ \mathrm{s/eV}\ ,
\end{equation}
is set by the nonobservation of infrared (IR) photons that would
be emitted by Big Bang relic neutrinos \cite{Ress} and, in
addition, by the nonobservation of scattering of high energy
[$O$(TeV)] photons from distant sources on this hypothetical IR
background \cite{Bill}. In the context of supernova neutrinos, the
lack of excess $\gamma$ flux during the SN 1987A neutrino burst
sets a limit $\tau/m\gtrsim O(10^{15})$~s/eV \cite{KoTu,Raft}. In
this Appendix, we estimate phenomenological limits on $\tau/m$
from SRN radiative decays which, to our knowledge, have not been
discussed so far in the literature.

Hypothetical radiative decays of SRN produce a diffuse photon
background, whose local $(z=0)$ density per unit of volume and
energy is obtained by integrating over the (redshifted) source
terms $q^\gamma_{ji}$,
\begin{equation}
\label{ngamma}
n_\gamma(E_\gamma)=\int_0^\infty
\frac{dz}{H(z)}\sum_{ji}^{m_j>m_i} q_{ji}^\gamma(E_\gamma(1+z),z)
\end{equation}
where the $q^\gamma_{ji}$ have an expression analogous to
Eq.~(\ref{qji}), but with the appropriate final-state energy
distribution for the photon in the laboratory frame,
\begin{equation}
\label{psigamma}
\psi^\gamma_{\nu_j\to\nu_i}(E_j,E_\gamma)=\frac{m^2_j}{E_j\,\Delta
m^2_{ji}} \left(1-\alpha+2\alpha\frac{E_\gamma\,m^2_j}{E_j\Delta
m^2_{ji}} \right)\
\end{equation}
(for $E_j>E_\gamma \Delta m^2_{ji}/m^2_j$), where $\Delta
m^2_{ji}=m^2_j-m^2_i$, and the so-called asymmetry (or anisotropy)
parameter $\alpha$ quantifies the amount of parity violation in
the decay \cite{Raft,Wilc} ($\alpha\in[-1,+1]$ for Dirac
neutrinos, while $\alpha=0$ for Majorana neutrinos \cite{Raft}).
In calculating the photon density%
\footnote{For photons ($v=c$), the density $n_\gamma$ (number  of
$\gamma$ per unit of volume and energy) also represents the flux
(number of $\gamma$ per unit of area, time, and energy) in natural
units.}
$n_\gamma$ from Eqs.~(\ref{ngamma}) and (\ref{qji}), one can
assume at first order that the neutrino density at any $z$ is
basically ``undecayed,'' namely
\begin{equation}
\label{nu0} n_{\nu_j}(E,z)\simeq
n^0_{\nu_j}(E,z)=\frac{1}{1+z}\int_0^\infty\frac{dz'}{H(z')}R_\mathrm{SN}(z')Y_{\nu_j}
\left(E\frac{1+z'}{1+z}\right)\ ,
\end{equation}
as derived from Eq.~(\ref{Solution}) in the limit $\Gamma_j
m_j/E_j\to 0$. This assumption will be validated {\em a
posteriori\/} below.

In order to perform numerical calculations, the neutrino mass and
decay parameters must be fixed. We assume a representative NH
scenario with $m_1=0$, so that  $m^2_j/ \Delta m^2_{ji}\simeq 1$
(strong hierarchy approximation) and the $\psi$ function in
Eq.~(\ref{psigamma}) becomes the same for all decay channels. We
also assume that $\tau_2/m_2=\tau_1/m_1\equiv\tau/m$, so that the
total photon flux is
\begin{equation}
\label{ngamma2} n_\gamma(E_\gamma)=(\tau/m)^{-1}\int_0^\infty
\frac{dz}{H(z)}\int_{E_\gamma}^\infty
\frac{dE'}{E'}\psi(E',E_\gamma) \sum_{j=2,3}^{} n^0_{\nu_j}(E',z)\
,
\end{equation}
independently of the decay branching ratios (provided that they
add up to unity, i.e., that only radiative decays occur). In the
above equation, the inner sum extends to all the unstable states
(both $\nu$ and $\nubar$). From Table~I, it follows that the
relevant neutrino yield to be integrated is
\begin{equation}
\label{yieldgamma} Y_{\nu_2}+Y_{\nu_3}+ Y_{\nubar_2}+Y_{\nubar_3}=
Y_{\nu_e}+3Y_{\nu_x}\ ,
\end{equation}
for any value of $P_H$.

Figure~7 shows the results for the photon flux
$n_\gamma(E_\gamma)$ in the above scenario, assuming
$\tau/m=10^{14}$~s/eV and  three representative values
$\alpha=-1,0,+1$, corresponding to green, red, and blue curves,
respectively. The curves cross at $E_\gamma\simeq 1.21$~MeV, where
the $\alpha$-dependent part of the integral in Eq.~(\ref{ngamma2})
vanishes. Lower (higher) values of $\tau/m$ would simply shift the
curves upwards (downwards) along the $y$-axis. In the same figure,
we superpose (as a dashed line) the best fit to the experimental
$\gamma$ background flux measured by the COMPTEL experiment for
$E_\gamma \in [0.1,\,30]$~MeV \cite{COMT},
\begin{equation}
n_\gamma^{\mathrm{bkgd}}(E_\gamma)\simeq 4\pi \times (1.05\pm
0.2)\times 10^{-4}\left(\frac{E_\gamma}{5\
\mathrm{MeV}}\right)^{-2.4}\mathrm{\ cm}^{-2}\mathrm{\
s}^{-1}\mathrm{\ MeV}^{-1}\ .
\end{equation}

From Fig.~7 we derive that, in order to keep the $\gamma$ flux
from hypothetical SRN decay below the observed $\gamma$
background, it must be
\begin{equation}\label{SRNlimit}
  \tau/m\gtrsim O(10^{14})\mathrm{\ s/eV}\ ,
\end{equation}
up to an $\alpha$-dependent factor of $O(1)$. Similar results
would
be obtained for inverted hierarchy and $m_3=0$ (not shown).%
\footnote{Radiative decays of quasidegenerate neutrinos would
instead produce $\gamma$'s with typical energy much lower than in
Fig.~7 for both NH and IH (not shown). We have checked that the
corresponding bound on $\tau/m$ would be degraded by roughly two
orders of magnitude, as compared with Eq.~(\ref{SRNlimit}).}
Notice that this limit largely justifies {\em a posteriori\/} the
``undecayed'' approximation for $n_\nu$ in Eq.~(\ref{nu0}).

We do not further elaborate the SRN constraint in Eq.~(\ref{nu0}),
since it is not competitive with the one in
Eq.~(\ref{stronglimit}). We note, however, that the supernova
relic neutrino bound in Eq.~(\ref{SRNlimit}) is stronger than the
one coming from the diffuse extragalactic stellar neutrino
background [$\tau/m\gtrsim O( 10^{12})$~s/eV] recently discussed
in \cite{Fior}.

\begin{figure}[t]
\vspace*{-0.0cm}\hspace*{-0.5cm}
\includegraphics[scale=0.6]{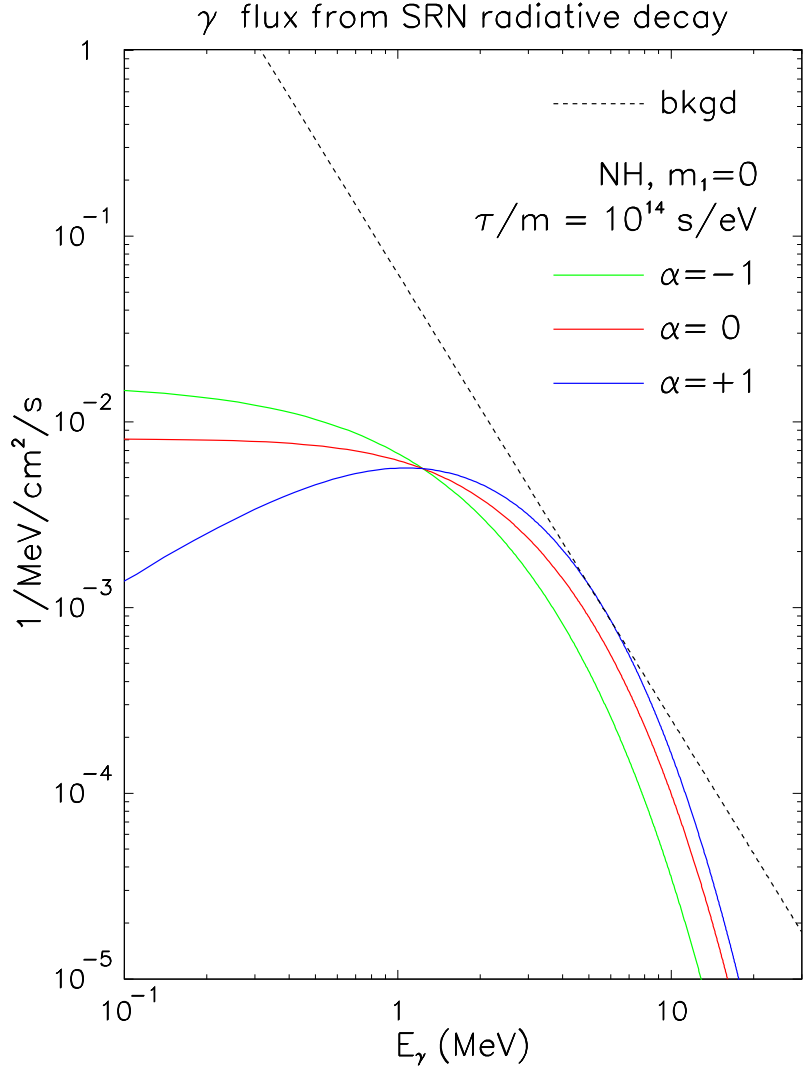}
\vspace*{+0.0cm} \caption{\label{fig7}
\footnotesize\baselineskip=4mm (Color online.) Energy spectrum of
the photon flux coming from hypothetical radiative decays of SRN
(solid curves), in a scenario with normal hierarchy, $m_1=0$, and
$\tau_2/m_2=\tau_3/m_3=10^{14}$~s/eV. The calculations are
performed for three representative values of the decay asymmetry
parameter $\alpha$ ($=-1,0,+1$). The dashed line represents the
power-law best fit to the $\gamma$ background measured by the
COMPTEL experiment in the range $E_\gamma \in [0.1,30]$~MeV
\cite{COMT}.}
\end{figure}

\section{Remarks on $3\nu$ oscillations of the atmospheric
neutrino background}

In water-Cherenkov detectors, $3\nu$ flavor transitions affect not
only the $e^+$ signal induced by supernova relic $\bar\nu_e$, but
also the $e^\pm$ background induced by low-energy atmospheric
neutrinos. The irreducible components of the $e^\pm$ background
\cite{SKSN,Male} are due to (1) interactions of atmospheric
$\nu_e$ and $\bar\nu_e$; and (2) decays of ``invisible'' $\mu^\pm$
(below the threshold for Cherenkov emission), induced by
low-energy atmospheric $\nu_\mu$ and $\bar\nu_\mu$. In
Super-Kamiokande \cite{SKSN}, the background components (1) and
(2) are characterized by parent neutrino energy spectra  peaked at
$E\sim 100$ and $E\sim 150$ MeV, respectively \cite{Male}.

In \cite{SKSN}, the effects of neutrinos oscillations on the
$e^\pm$ background are calculated by assuming pure $2\nu$
oscillations in the $\nu_\mu\to\nu_\tau$ channel with maximal
mixing (i.e., $\delta m^2\simeq 0$, $\theta_{13}\simeq 0$, and
$\sin^2\theta_{23}\simeq 1/2$).  Within the $2\nu$ approximation,
the background component (1) is not affected by oscillations,
while the component (2) is suppressed by the average effect of
$\Delta m^2$-driven oscillations, through a factor $f_\mu=\sin^2
2\theta_{23}\sin^2(1.27 \Delta m^2 L/E)\simeq 0.5\sin^2
2\theta_{23}\simeq 0.5$, where $[\Delta m^2]=\mathrm{eV}^2$,
$[L]=\mathrm{m}$, and $[E]=\mathrm{MeV}$. In terms of averaged
flavor oscillation probabilities $P_{\alpha\beta}$, the $2\nu$
approximation implies that $(P_{ee}, P_{\mu\mu}, P_{\mu e})\simeq
(1,\,0.5,\,0)$ \cite{SKSN}.

We remark, however, that the $2\nu$ approximation $\delta
m^2\simeq 0$, usually valid for typical atmospheric neutrino
energies ($\gtrsim 1 $~GeV), is not applicable in the SRN context.
Indeed, for $E\sim O(100)$~MeV and $L\sim O(R_\oplus)$, the
associated oscillation phase cannot be neglected [$\delta m^2
L/E\sim O(1)$]. Even for $\theta_{13}=0$, the low-energy
atmospheric neutrino background is thus sensitive to the ``solar''
$\delta m^2$, rather than to the ``atmospheric'' $\Delta m^2$
(which can be effectively taken as $\infty$).

The possibility of probing $\delta m^2$-driven oscillations at
very low atmospheric $\nu$ energy ($\ll 1$~GeV) would deserve, in
itself, a
separate investigation.%
\footnote{For $E\gtrsim O(1)$~GeV, subleading oscillations driven
by $\delta m^2$ have instead been considered in several papers;
see, e.g., \cite{Pere} and references therein.}
Here we only make some rough estimates of the relevant flavor
oscillation probabilities $P_{\alpha\beta}$ for $\delta m^2\neq
0$.  We assume isotropic atmospheric neutrino fluxes (produced at
$h=20$~km from the ground level), equal components of $\nu$ and
$\bar\nu$ (with weights 3 and 1, respectively, to account for the
different cross sections), and a $\nu_\mu/\nu_e$ flavor ratio
$r=2$. Matter effects are estimated through a constant-density
approximation for the Earth. The oscillations parameters are taken
from Eqs.~(\ref{dm2})--(\ref{s23}), with either
$\sin^2\theta_{13}=0$ or $\sin^2\theta_{13}=0.067$ [see
Eq.~(\ref{s13})]. We consider a representative neutrino energy
$E=100$~MeV, and average over all incoming $\nu$ directions.

Under these assumptions, and for $\sin^2\theta_{13}=0$, we find
$(P_{ee}, P_{\mu\mu}, P_{\mu e})\simeq (0.77,\,0.42,\,0.11)$,
which differ significantly from the values $(1,\,0.5,\,0)$
obtained by setting $\delta m^2=0$. The atmospheric electron
events [background (1)] and invisible muon events [background (2)]
are then modulated, respectively, by the factors $f_e\simeq
P_{ee}+rP_{\mu e}\simeq 0.99$ and $f_\mu\simeq
P_{\mu\mu}+P_{ee}/r\simeq
0.48$, which happen to differ only slightly%
\footnote{An accidental cancellation of effects is operative for
the specific value $r=2$. See, e.g., \cite{Pere}.}
 from the values $(f_e,\,f_\mu)=(1,\,0.5)$ considered in
\cite{SKSN} by setting $\delta m^2=0$. Analogously, for
$\sin^2\theta_{13}=0.067$ we find $(P_{ee}, P_{\mu\mu}, P_{\mu
e})\simeq (0.67,\,0.17,\,0.41)$ and $(f_e,\,f_\mu)\simeq
(1.01,\,0.50)$. Therefore, despite significant $\delta
m^2$-induced effects on the oscillation probabilities, the
irreducible atmospheric background rates do not appear to be
significantly modified as compared with those obtained by
(incorrectly) setting $\delta m^2=0$, independently of the value
of $\sin^2\theta_{13}$.

In conclusion, the approximation $\delta m^2=0$ is not applicable,
in principle, to the analysis of the irreducible atmospheric
neutrino background in SRN searches \cite{SKSN}. However, in
practice, this approximation does not appear to be harmful
(according to our provisional estimates). More refined estimates
of $\delta m^2$-driven oscillations might be of some interest in
future high-statistics SRN searches \cite{UNNO,Hype}, which are
expected to find a signal above the irreducible atmospheric $\nu$
background. These oscillation effects would instead loose interest
if the ``irreducible'' background could  be ``reduced away''  by
tagging SRN events, as recently proposed in \cite{GADZ}.



\begin{thebibliography}{99}

\bibitem{Revi}  V.~Barger, D.~Marfatia and K.~Whisnant,
                Int.\ J.\ Mod.\ Phys.\ E {\bf 12}, 569 (2003);
                M.~C.~Gonzalez-Garcia and Y.~Nir,
                Rev.\ Mod.\ Phys.\  {\bf 75}, 345 (2003);
                C.~Giunti and M.~Laveder, hep-ph/0310238;
                G.~Altarelli and K.~Winter (editors),
                {\em Neutrino Mass}, Springer Tracts in Modern Physics {\bf 190} (2003).

\bibitem{Hagi}  K.~Hagiwara {\it et al.}
                [Particle Data Group Collaboration],
                Phys.\ Rev.\ D {\bf 66}, 010001 (2002).

\bibitem{Sola}  G.L.~Fogli, E.~Lisi, A.~Marrone, and A.~Palazzo,
                hep-ph/0309100, to appear in Phys.\ Lett.\ B.

\bibitem{Atmo}  G.L.~Fogli, E.~Lisi, A.~Marrone, D.~Montanino,  A.~Palazzo, and A.M.~Rotunno,
                Phys.\ Rev.\ D {\bf 69}, 017301 (2004).

\bibitem{Dolg}  A.D.~Dolgov,    Phys.\ Rept.\  {\bf 370}, 333 (2002).

\bibitem{Pakv}  S.~Pakvasa,
                hep-ph/0305317, in the Proceedings of the 10th International
                Workshop on Neutrino Telescopes (Venice, Italy, 2003), edited by
                M.~Baldo-Ceolin (U.\ of Padua Publication, Italy, 2003), p.~469.

\bibitem{Gelm}  G.~Gelmini and E.~Roulet,
                Rept.\ Prog.\ Phys.\  {\bf 58}, 1207 (1995).

\bibitem{Pak2}  S.~Pakvasa, hep-ph/9905426,
                in the Proceedings of the 8th International
                Workshop on Neutrino Telescopes (Venice, Italy, 1999), edited by
                M.~Baldo-Ceolin (U.\ of Padua Publication, Italy, 1999), p.~283.

\bibitem{Bell}  J.F.~Beacom and N.F.~Bell,
                Phys.\ Rev.\ D {\bf 65}, 113009 (2002);
                see also A.~Bandyopadhyay, S.~Choubey, and S.~Goswami,
                Phys.\ Lett.\ B {\bf 555}, 33 (2003).

\bibitem{KaNu}  KamLAND Collaboration,
                K.~Eguchi {\em et al.}, hep-ex/0310047.


\bibitem{SN87}  J.A.~Frieman, H.E.~Haber, and K.~Freese,
                Phys.\ Lett.\ B {\bf 200}, 115 (1988).

\bibitem{Be03}  J.F.~Beacom, N.F.~Bell, D.~Hooper, S.~Pakvasa, and T.J.~Weiler,
                Phys.\ Rev.\ Lett.\  {\bf 90}, 181301 (2003);
                G.~Barenboim and C.~Quigg, Phys.\ Rev.\ D {\bf 67}, 073024 (2003).

\bibitem{SKSN}  Super-Kamiokande Collaboration, M.~Malek {\it et al.},
                Phys.\ Rev.\ Lett.\  {\bf 90}, 061101 (2003).

\bibitem{Male}  M.S.~Malek, PhD thesis (SUNY, Stony Brook, 2003),
                available at www-sk.icrr.u-tokyo.ac.jp/sk/pub/index.html~.

\bibitem{Ando}  S.~Ando, Phys.\ Lett.\ B {\bf 570}, 11 (2003).

\bibitem{GADZ}  J.F.~Beacom and M.R.~Vagins, hep-ph/0309300,
                submitted to Phys.\ Rev.\ Lett.

\bibitem{UNNO}  C.K.~Jung, hep-ex/0005046. See also the site:
                superk.physics.sunysb.edu/nngroup/uno~.

\bibitem{Hype}  K.~Nakamura,
                Int.\ J.\ Mod.\ Phys.\ A {\bf 18}, 4053 (2003).

\bibitem{An03}  S.~Ando, K.~Sato, and T.~Totani,
                Astropart.\ Phys.\  {\bf 18}, 307 (2003).

\bibitem{AnSa}  S.~Ando and K.~Sato,
                Phys.\ Lett.\ B {\bf 559}, 113 (2003).

\bibitem{Fuku}  M.~Fukugita and M.~Kawasaki,
                Mon.\ Not.\ Roy.\ Astron.\ Soc.\ {\bf 340}, L7 (2003).

\bibitem{Ka03}  L.E.~Strigari, M.~Kaplinghat, G.~Steigman, and T.P.~Walker,
                astro-ph/0312346.

\bibitem{An04}  S.~Ando, astro-ph/0401531.

\bibitem{Mada}  P.~Madau {\em et al.},
                Mon.\ Not.\ Roy.\ Astron.\ Soc.\  {\bf 283}, 1388 (1996);
                P.~Madau, M.~Della~Valle, and N.~Panagia,
                Mon.\ Not.\ Roy.\ Astron.\ Soc.\ {\bf 297}, 17 (1998).

\bibitem{SDSS}  See, e.g., SDSS Collaboration, M.~Tegmark {\it et al.},
                astro-ph/0310723.

\bibitem{SNan}  G.L.~Fogli, E.~Lisi, D.~Montanino, and A.~Palazzo,
                Phys.\ Rev.\ D  {\bf 65}, 073008 (2002)
                [Erratum-ibid.\ D {\bf 66}, 039901 (2002)].

\bibitem{Matt}  L.~Wolfenstein, Phys.\ Rev.\ D {\bf 17}, 2369 (1978);
                S.P.~Mikheev and A.Yu.\ Smirnov, Yad.\ Fiz.\ {\bf 42}, 1441 (1985)
                [Sov.\ J.\ Nucl.\ Phys.\ {\bf 42}, 913 (1985)].

\bibitem{Shoc}  R.C.~Schirato and G.M.~Fuller, astro-ph/0205390;
                G.L.~Fogli, E.~Lisi, D.~Montanino, and A.~Mirizzi,
                Phys.\ Rev.\ D {\bf 68}, 033005 (2003).

\bibitem{KuPa}  T.K.~Kuo and J.~Pantaleone,
                Rev.\ Mod.\ Phys.\ {\bf 61}, 937 (1989).

\bibitem{Luna}  C.~Lunardini and A.Yu.~Smirnov,
                JCAP {\bf 0306}, 009 (2003).

\bibitem{Spec}  G.G.~Raffelt, M.T.~Keil, R.~Buras, H.T.~Janka, and M.~Rampp,
                astro-ph/0303226, to appear in the Proceedings of {\em NOON 2003},
                4th International Workshop on Neutrino Oscillations and their
                Origin (Kanazawa, Japan, 2003); M.T.~Keil, G.G.~Raffelt, and
                H.T.~Janka, Astrophys.\ J.\  {\bf 590}, 971 (2003).

\bibitem{Xsec}  P.~Vogel, Prog.\ Part.\ Nucl.\ Phys.\ {\bf 48}, 29 (2002);
                P.~Vogel and J.~F.~Beacom, Phys.\ Rev.\ D {\bf 60}, 053003 (1999).

\bibitem{Lind}  M.~Lindner, T.~Ohlsson and W.~Winter,
                Nucl.\ Phys.\ B {\bf 607}, 326 (2001).

\bibitem{Bere}  Z.G.~Berezhiani and M.I.~Vysotsky,
                Phys.\ Lett.\ B {\bf 199}, 281 (1987);
                Z.G.~Berezhiani, G.\ Fiorentini, M.~Moretti, and A.~Rossi,
                Z.\ Phys.\ C {\bf 54}, 581 (1992).

\bibitem{GLam}  C.~Giunti, C.~W.~Kim, U.W.~Lee, and W.P.~Lam,
                Phys.\ Rev.\ D {\bf 45}, 1557 (1992).

\bibitem{Kach}  M.~Kachelriess, R.~Tom{\`a}s, and J.W.F.~Valle,
                Phys.\ Rev.\ D {\bf 62}, 023004 (2000).

\bibitem{Smir}  Z.G.~Berezhiani and A.Yu.~Smirnov,
                Phys.\ Lett.\ B {\bf 220}, 279 (1989).

\bibitem{Choi}  K.~Choi and A.~Santamaria,
                Phys.\ Rev.\ D {\bf 42}, 293 (1990).

\bibitem{Toma}  R.~Tom{\`a}s, H.~P{\"a}s, and J.W.F.~Valle,
                Phys.\ Rev.\ D {\bf 64}, 095005 (2001).

\bibitem{Farz}  Y.~Farzan,
                Phys.\ Rev.\ D {\bf 67}, 073015 (2003).

\bibitem{Kolb}  E.W.~Kolb and M.S.~Turner,
                {\em The Early Universe\/} (Westview Press, Boulder, 1994).

\bibitem{Pecc}  Y.\ Chikashige, R.~Mohapatra, and R.~Peccei,
                Phys.\ Lett.\ B {\bf 98}, 265 (1981);
                G.~Gelmini and M.~Roncadelli, Phys.\ Lett.\ B
                {\bf 99}, 411 (1981).

\bibitem{PaVa}  S.~Pakvasa and J.W.F.~Valle, hep-ph/0301061.

\bibitem{Raft}  G.G.~Raffelt, {\em Stars as Laboratories for Fundamental
                Physics\/} (U.\ of Chicago Press, Chicago, 1996).

\bibitem{KLam}  C.W.~Kim and W.P.~Lam,  Mod.\ Phys.\ Lett.\ A {\bf 5}, 297 (1990).

\bibitem{El74}  S.~Eliezer and D.A.~Ross,
                Phys.\ Rev.\ D {\bf 10}, 3088 (1974).

\bibitem{Pe77}  S.T.~Petcov,
                Sov.\ J.\ Nucl.\ Phys.\ {\bf 25}, 340 (1977).

\bibitem{Ress}  M.T.~Ressel and M.S.~Turner,
                Comments on Astrophys.\ {\bf 14}, 323 (1990).


\bibitem{Bill}  S.D.~Biller {\em et al.},
                Phys.\ Rev.\ Lett.\ {\bf 80}, 2992 (1998);
                see also the comment by G.G.\ Raffelt,
                Phys.\ Rev.\ Lett.\ {\bf 81}, 4020 (1998).

\bibitem{KoTu}  E.W.\ Kolb and M.S.\ Turner,
                Phys.\ Rev.\ Lett.\ {\bf 62}, 509 (1989).

\bibitem{Wilc}  L.F.\ Li and F.\ Wilczek,
                Phys.\ Rev.\ D {\bf 25}, 143 (1982).

\bibitem{COMT}  COMPTEL Collaboration,
                AIP Conf.\ Proc.\ {\bf 510}, 392 (2000); S.C.~Kappadath,
                PhD thesis (U.\ of New Hampshire, 1998), available at
                wwwgro.unh.edu/users/ckappada~.

\bibitem{Fior}  C.\ Porciani, S.\ Pietroni, and G.\ Fiorentini,
                astro-ph/0311489.

\bibitem{Pere}  O.L.G.~Peres and A.Yu.~Smirnov, hep-ph/0309312.









\end{thebibliography}
\end{document}